\documentclass[12pt,aps,floats,showpacs,amssymb,tightenlines]{revtex4}

\usepackage{color}
\usepackage{graphicx}
\usepackage{amsmath}
\usepackage{amsfonts}
\usepackage{amscd}
\usepackage{epsfig}
\usepackage{amssymb}
\usepackage{tabularx}

\begin{document}

\title{On the evolution of the momentarily static radiation free data in the Apostolatos - Thorne cylindrical shell model.}
\author{Reinaldo J. Gleiser} \email{gleiser@fis.uncor.edu} \author{Daniel E. Barraco}

\affiliation{Instituto de F\'{\i}sica Enrique Gaviola and FAMAF,
Universidad Nacional de C\'ordoba, Ciudad Universitaria, (5000)
C\'ordoba, Argentina}

\begin{abstract}
In this paper we study the evolution of the ``Momentarily Static and Radiation Free'' (MSRF) initial data for the Apostolatos - Thorne cylindrical shell model. After briefly reviewing the equations of motion, the definition of the  MSRF initial data and of its relation to the static solution that corresponds to the given conserved intrinsic parameters of the shell, we show that for MSRF data the initial acceleration of the shell is always directed towards the static radius. We analyze in detail the relation between the parameters characterizing the configuration corresponding to the initial data and those for the assumed final static configuration, and show that, once the appropriate properties of the solutions of the cylindrical wave equation are taken into account, there is a priori no conflict for any choice of initial MSRF data, in contrast with some recent results of Nakao, Ida and Kurita. To obtain a more detailed description of the evolution we consider the case where the problem can be analyzed in the linear approximation, and show that the evolution is stable in all cases. The possible form of the approach to the final static configuration is also analyzed. We find that this approach is very slow, with an inverse logarithmic dependence on time at fixed radius. Given the absence of analytic solutions for the problem, we introduce a numerical computation procedure that allows us to visualize the explicit form of the evolution of the shell and the gravitational field up to large times. The results are in agreement with the qualitative behaviour conjectured by Apostolatos and Thorne, with an initial damped oscillatory stage, but we find that this oscillations are not about the final static radius but rather about a position that approaches slowly that of the static final state, as indicated by our analysis. We also include an Appendix, where we review some properties of the solutions of the cylindrical wave equation, and prove the existence of solutions with vanishing initial value for $r > R_0$, ($R_0 > 0$ some finite constant), that approach a constant value for large times. This result is crucial for the proof of compatibility of arbitrary MSRF initial data and a final static configuration for the system.

\end{abstract}


\maketitle
\section{Introduction}
The Apostolatos-Thorne model \cite{apostol} describes the dynamics
of a self gravitating cylindrical shell of counter rotating
particles. It was originally introduced in the context of a discussion of the effect of rotation on halting the collapse of the system. But the system is interesting also on other grounds, because it provides a model of a mechanical system, with a simple Newtonian limit, in interaction with a dynamical gravitational field, interchanging energy with the gravitational radiation contents of the field. As shown in \cite{apostol}, it is not difficult to obtain a set of coupled ordinary and partial derivative equations for the dynamical variables whose solutions describe the possible evolutions of the system. Although the original paper \cite{apostol} did not give a detailed computation of the evolution of the system, it included an extensive discussion of the main qualitative features that would result from the interaction of the shell with the gravitational field. There has recently been a renewed interest in finding solutions and analyzing this detailed evolution in several cases and for different types of initial data. In \cite{hamity}, Hamity, C\'ecere and Barraco considered the evolution of the outer shell in a system that contains also an inner shell to allow for the imposition of a particular boundary condition. They also make use of some simplifying assumptions that reduce the full set of evolution equations to a particular ordinary differential equation for the motion of the shell. A different approach, where the evolution of the system is analyzed in the linear approximation was considered by Gleiser and Ramirez in \cite{gleram}.  The stability of the static configurations of the system under arbitrary perturbations was proved in an extended analysis given in \cite{gleramqnm}, where the existence of quasi normal modes was also considered. A related analysis, but with a different interpretation was also given in \cite{nakaoqnm}.

A particular case considered in \cite{apostol} was that resulting from initial data with a structure similar to that corresponding to a static solution, but differing from that one, so that the shell starts its motion with a non vanishing acceleration. This type of initial data was designated ``Momentarily Static and Radiation Free'' (MSRF), and the authors of \cite{apostol} considered it as both an example of consistent initial data for the system and as an example of the possible evolution resulting from the dynamic equations. No explicit solution is given in \cite{apostol}, but the authors analyze explicitly the initial acceleration of the shell and provide qualitative arguments for the resulting evolution, concluding that the shell would execute damped oscillations about a final static radius, as the mechanical energy of the shell is transferred to the gravitational field, and radiated away. This conclusion, however, has been challenged in a recent paper by Nakao and Kurita \cite{nakao}. In this paper the authors analyze some constraints that the Einstein equations for the system would impose on the possible final state of the evolution, starting with MSRF, and conclude that a static solution could not be reached from a certain set of MSRF data. This result would imply, for instance, that if the system started with that type of MSRF data then it could not reach a static configuration, and would expand forever, something that could not happen in the Newtonian limit, and therefore, we would be confronted with a new type of instability, resulting from the interaction of the shell with the dynamical part of the gravitational field.
This, if correct, would indeed be a very surprising and unexpected feature of the dynamics of this system, since it has a simple, well defined Newtonian limit, where {\em all} motions are bounded, as a result of the corresponding Newtonian potential being unbounded as one moves further and further away from the symmetry axis, and it is not easy to understand in what way the radiative modes could modify this feature so drastically. It is for that reason that we considered important to provide a different analysis of the evolution of MSRF data, in order to either confirm the results of \cite{nakao}, and try to find a possible interpretation, or on the contrary, provide arguments, and possible proofs, in favor of the original qualitative view of that evolution, as discussed in \cite{apostol}. As we shall show, although there are no errors in \cite{nakao}, their result is a consequence of a particular assumption on the behaviour of the gravitational field at late times that needs to be revised. Once this is done we find no conflict between an arbitrary MSRF initial data and a corresponding final static configuration, confirming the conjecture in \cite{apostol}. Furthermore, we consider MSRF data that is initially close to the static configuration, and, using some results obtained in \cite{gleramqnm}, prove the stability of the evolution. We further apply a numerical integration procedure and obtain the detailed evolution of the system up to large times, finding somewhat unexpected features in the approach of the shell to its final static radius, that are described and discussed in the text below.

The plan of the paper is as follows. In the next Section we review the formulation of the Apostolatos - Thorne model and of its equations of motion. In Section III, we recall the definition of MSRF initial data, and its relation to the conserved intrinsic parameters of the shell. Next we find the static solution that corresponds to the same set of intrinsic parameters, and find the relation between the initial state free parameters and those of the final state. This leads to a relation between the initial and final forms of one of the metric functions that was claimed in \cite{nakao} to lead in some cases to a contradiction with the equations of motion. However, as we show in the Appendix, this result is a consequence of an inadequate use of a general expression for the solution of the wave equation with cylindrical symmetry, and we prove the existence of solutions of the field equations that satisfy the relations that make {\em all} initial MSRF data compatible with a static final state. In Section IV we consider small departures from the static configuration, in the sense that the initial radius is close to the corresponding static radius, analyze the resulting linearized equations of motion, and, making use of a result obtained in \cite{gleramqnm}, prove the absolute stability of the evolution of MSRF data close to the static solution. We also show that, contrary to a simple expectation, the final approach to the static configuration, for fix radial coordinate, is very slow, with the appropriate quantities approaching zero only as $1/\ln(\tau)$, where $\tau$ is the proper time on the shell. Since the previous derivations give no information on the detailed evolution of the shell and the fields, in Section V we use a numerical integration procedure, developed in \cite{gleramqnm}, to visualize, for some particular examples, the evolution from the initial state, up to large times. We find for the shell, as expected, an initial exponentially damped oscillatory stage, but, in agreement it our analysis in Section IV, not about the final static radius, but rather about a position that approaches the static radius as $1/\ln(\tau)$.
We end the paper with some comments and conclusions, as well as comments on related work by other authors, in particular to that of reference \cite{nakao}.

\section{The Apostolatos - Thorne model}

The Apostolatos - Thorne model \cite{apostol} describes the dynamics
of a self gravitating cylindrical shell of counter rotating
particles. Both the inner ($M^-$) and  outer ($M^+$) regions of the shell are vacuum
space times with a common boundary $\Sigma$. The corresponding metrics may be written in the form,
\begin{equation}\label{ATeq1}
ds^2_{\pm} = e^{2\gamma_{\pm}-2 \psi_{\pm}}
\left(dr^2-dt^2_{\pm}\right)+ e^{2 \psi_{\pm}}dz^2 +e^{-2
\psi_{\pm}}r^2 d\phi^2
\end{equation}
where the $(+)$ sign corresponds to the outer, and $(-)$ to the
inner regions. The functions $\psi$, and $\gamma$ depend only on $r,t$
and satisfy the equations:
\begin{equation}\label{ATeq2}
    \psi_{,rr} +\frac{1}{r}\psi_{,r} -\psi_{,tt} = 0
\end{equation}
\begin{equation}\label{ATeq3}
\gamma_{,t} = 2 r \psi_{,r} \psi_{,t}\;\;\; , \;\;\; \gamma_{,r} = r
\left[(\psi_{,r})^2+(\psi_{,t})^2\right]
\end{equation}
The shell is located on the hypersurface $\Sigma$ given by $r=R(\tau)$,
where $\tau$ is the proper time of an observer at rest on the shell.
We may interpret $\psi(r, t)$ as playing the role of a gravitational
field whose static part is the analogue of the Newtonian potential.
The time dependent solutions of (\ref{ATeq2}) represent
gravitational waves (Einstein-Rosen). Equation (\ref{ATeq2}) is the
integrability condition for Eqs. (\ref{ATeq3}). The coordinates $(z,
\phi, r )$ and the metric function $\psi$ are continuous across the
shell $\Sigma$, while $t$ and the metric function $\gamma$ are
discontinuous. Smoothness of the spacetime geometry on the axis $r =
0$ requires that $ \gamma = 0$, $\psi$ finite at $r = 0$, and $\partial \psi/\partial r|_{r=0} =0$. The
junction conditions of $M^-$ and $M^+$ through $\Sigma$ require the
continuity of the metric and specify the jump of the extrinsic
curvature $K^{\pm}$ compatible with the stress energy tensor on the
shell. The induced metric on $\Sigma$ is given by
 \begin{equation}\label{ATeq4}
ds^2_{\Sigma} =-d \tau^2 + e^{2 \psi_{\Sigma}} dz^2 +  e^{-2
\psi_{\Sigma}} R^2 d\phi^2
 \end{equation}
where $\psi_{\Sigma}(\tau) =
\psi_+(R(\tau),t_+(\tau))=\psi_-(R(\tau),t_-(\tau))$.

The evolution
of the shell is characterized by $R(\tau)$. If we assume that the shell is
made up of equal mass counter rotating particles, the Einstein field
equations on the shell may be put in the form,
\begin{equation}\label{ATeq5}
\psi^+_{,n}-\psi^-_{,n} = -\frac{2 \lambda} {  \sqrt{R^2+
e^{2\psi_{\Sigma}} J^2}  }
\end{equation}
\begin{equation}\label{ATeq6}
X^+-X^- = -\frac{4 \lambda \sqrt{R^2+e^{2 \psi_{\Sigma}}J^2}}{R}
\end{equation}
where the constants $\lambda$ and $J$ are, respectively, the proper
mass per unit Killing length of the cylinder and the angular
momentum per unit mass of the particles. The other quantities in
(\ref{ATeq5},\ref{ATeq6}) are given by,
\begin{equation}\label{ATeq7}
X^{\pm} \equiv \frac{\partial t_{\pm}}{\partial \tau} =
+\sqrt{e^{-2(\gamma_{\pm}-\psi_{\Sigma})} +\dot{R}^2}
\end{equation}
\begin{equation}\label{ATeq8}
\psi^{\pm}_{,n} = \psi^{\pm}_{,r} X^{\pm} +\psi^{\pm}_{,t} \dot{R}
\end{equation}
where a dot indicates a $\tau$ derivative, and we also have,
\begin{eqnarray}
\label{ATeq9} \frac{d^2 R}{d\tau^2} &=&  \dot{R}\dot{\psi_{\Sigma}}
- R
\left[(\dot{\psi_{\Sigma}})^2 +(\psi^-_{,n})^2\right]  \nonumber \\
& & +\frac{ R^2 \psi^-_{,n} X^-}{R^2+e^{2 \psi_{\Sigma}}J^2}
-\frac{\lambda R^2 X^-}{(R^2+e^{2 \psi_{\Sigma}}J^2)^{3/2}}
+\frac{J^2 e^{2 \psi_{\Sigma}}X^-X^+}{R(R^2+e^{2 \psi_{\Sigma}}J^2)}
\end{eqnarray}

These equations together with (\ref{ATeq2},\ref{ATeq3}) determine
the evolution of the shell and of the gravitational field to which
it is coupled.

\section{The momentarily static radiation free (MSRF) initial data.}

The set of equations of the previous Section may, at least in principle, be solved as an initial plus boundary (plus matching conditions) problem. Namely, we expect that given appropriate initial data, there should be a well defined evolution to the future of that data. An inspection of the equations indicate that such initial data could be specified on a space like hypersurface formed by taking a constant time $t_+=t_{0+}$ slice on $M^+$ and a constant time $t_-=t_{0-}$ slice on $M^-$, matched through a constant $\tau$ section of $\Sigma$. The independent data on $M^+$ would then be the functions $\psi_+(t_{0+},r)$ and $\partial \psi_+(t_{0+},r)/\partial t$, since these specify $\gamma_+(t_{0+},r)$ and $\partial \gamma_+(t_{0+},r)/\partial t$ up to a constant. Similarly, on $M^-$, we may give arbitrary expressions for $\psi_-(t_{0-},r)$ and $\partial \psi_-(t_{0-},r)/\partial t$, and $\gamma_-(t_{0-},r)$ and $\partial \gamma_-(t_{0-},r)/\partial t$ are then defined up to a constant. We need also to specify some initial data for the shell, which could the (independent) values of $R$ and $d R/ d \tau$, and fix some of the constant parameters such as $J$ and $\lambda$. All these data is constrained in part by the matching conditions on $\Sigma$. Notice also that if we require regularity on the axis $r=0$ we need to impose $\gamma_-(t_-,0)=0$, and $\partial \psi_-(t_-,0)/\partial r = 0$. In the rest of this paper we shall only consider data and evolutions satisfying these requirements. \\

A particular family of initial data for the model, recently considered in \cite{nakao}, was introduced by Apostolatos - Thorne in \cite{apostol}, and identified as the {\em momentarily static radiation free (MSRF) initial data}. It is defined as follows. We notice that $t_-$, $t_+$, and $\tau$ are defined up to arbitrary additive constants. We may therefore consider the points on $\Sigma$ corresponding to a given value of $\tau$, say $\tau=0$, and the corresponding hypersurfaces of constant $t_-$ and $t_+$ that are matched to $\Sigma$ at $\tau=0$, and assign also $t_-=0$ and $t_+=0$ to those hypersurfaces. Next we impose,
 \begin{equation}
  \label{equ01}
    \psi_+(0,r) = \psi_i-\kappa \ln(r/R_i)  \;\;\; ; \;\;\;  \gamma_+(0,r) = \gamma_i+\kappa^2 \ln(r/R_i)
  \end{equation}
 \begin{equation}
  \label{equ02}
  \left.\frac{\partial \psi_+(t_+,r)}{\partial t_+ }\right|_{t_+=0} = 0
  \end{equation}
   \begin{equation}
  \label{equ03}
   \psi_-(0,r) = \psi_i \;\;\; ; \;\;\; \gamma_-(0,r) = 0
 \end{equation}
  \begin{equation}
  \label{equ04}
   \left.\frac{\partial \psi_-(t_-,r)}{\partial t_- }\right|_{t_-=0} = 0
   \end{equation}
 \begin{equation}
  \label{equ05}
   R(0) = R_i \;\;\; ; \;\;\;
   \left.\frac{d R}{d \tau}\right|_{\tau=0} = 0
 \end{equation}

Notice that these conditions imply,
\begin{equation}
  \label{equ02a}
  \left.\frac{\partial \gamma_+(t_+,r)}{\partial t_+ }\right|_{t_+=0} = 0 \;\;\; ; \;\;\;
   \left.\frac{\partial \gamma_-(t_-,r)}{\partial t_- }\right|_{t_-=0} = 0
  \end{equation}

 A particular shell is described by fixed values of $\lambda$ and $J$. The matching conditions then impose constraints on the parameters that appear in (\ref{equ01} - \ref{equ05}). These may be written as,
\begin{equation}\label{equ06}
    2 \lambda R_i -\kappa e^{\psi_i-\gamma_i} \sqrt{R_i^2+e^{2\psi_i}J^2} =0,
\end{equation}

\begin{equation}\label{equ07}
   R_i e^{\psi_i-\gamma_i} -R_i  e^{\psi_i} + 4 \lambda \sqrt{R_i^2+e^{2\psi_i}J^2} =0
\end{equation}
and,
\begin{equation}\label{equ08}
\left. \frac{d^2 R}{d \tau^2} \right|_{\tau=0} + \frac{\lambda R_i^2 e^{\psi_i}}{\left(R_i^2+e^{2\psi_i} J^2\right)^{3/2}}
-\frac{ e^{4 \psi_i-\gamma_i}}{R_i \left(R_i^2+e^{2\psi}J^2\right)}=0
\end{equation}

The set (\ref{equ06} - \ref{equ08}) constrains the free parameters, and to a certain extent, their ranges. For instance, from (\ref{equ06}) and  (\ref{equ07}) we find,
\begin{equation}\label{equlam01}
    \lambda = \frac{\kappa e^{\psi_i} R_i\sqrt{R_i^2+e^{2 \psi_i} J^2} }{2(R_i^2+2 \kappa( R_i^2 +e^{2\psi_i} J^2))}
\end{equation}
and one can check that, for fixed $R_i$, $\kappa$, and $J$, $\lambda$ is a monotonic function of $\psi_i$ that satisfies the constraint $ 0 < \lambda < R_i/(4 J) $ for $-\infty < \psi_i < +\infty$.

In any case, there are many ways of handling these constraints. We may, for instance, assume that we fix $\lambda$, $J$, and the initial radius $R_i$. The value of $\psi_i$ could then be chosen freely, and we would use (\ref{equ06}) and (\ref{equ07}) to obtain the corresponding values of $\gamma_i$, and $\kappa$. Then Eq.(\ref{equ08}) determines the initial acceleration of the shell, and the initial data set is complete. In their original work, Apostolatos and Thorne noticed that in the Newtonian limit, a similar self gravitating shell, released from a momentarily static configuration, namely, with vanishing radial velocity, would undergo periodic oscillations about a particular static (``equilibrium'') configuration, fixed by its intrinsic parameters $J$ and $\lambda$. They therefore assumed that in the corresponding general relativistic problem, the shell would also execute oscillations about some ``equilibrium'' radius, but that now these oscillations would be damped as the system loses its mechanical energy through the emission of gravitational waves.

This conclusion, however, appears to be contradicted by the recent analysis by Nakao, et. al. \cite{nakao}, based on the properties of the evolution equations for the fields, that leads to the conclusion that for at least a subset of this type of initial data, contrary to the assumptions in \cite{apostol}, the system becomes unstable when the gravitational waves are taken into consideration, indicating, if correct, a surprising and rather drastic difference with the corresponding Newtonian problem. We need here to emphasize that no explicit example of the detailed evolution of the system under the MSRF initial data has been presented so far in the literature, and that the conclusions in \cite{nakao} were obtained under certain assumptions on the behaviour of the fields $\psi_+$ and $\gamma_+$ that, as we shall show, need not hold in the actual evolution of the system.

Going back to the system (\ref{equ06} - \ref{equ08}), and the evolution equations, we notice that if we further impose the condition $d^2 R/d\tau^2|_{\tau=0} = 0$, we find a solution of the evolution equations where $R(\tau)=R_i$, and the fields take the form (\ref{equ01} -\ref{equ04}) for {\em all} times, i.e., we have a {\em static} solution. (Hence the name MSRF for this type of initial data). Since, as indicated, $\psi_{\pm}$ are defined up to an arbitrary constant, we may choose $\psi_i=0$ for the static configuration.

Now, suppose that we have some MSRF initial data. If we assume that the evolution of this MSRF initial data leads (asymptotically) to a static configuration, can we find a relation between the MSRF data and the final static data? We recall that $\lambda$ and $J$ are constants of the motion, but there is also, for this type of initial data, another constant quantity, given by the coefficient $\kappa$. This constancy is analyzed in the Appendix, but here we proceed assuming that $\lambda$, $J$ and $\kappa$ are the same for the MSRF initial data and for the final static configuration. In particular, we assume that for the final state configuration we have,
\begin{equation}
  \label{equ01a}
    \psi_+(t_+,r) = \kappa \ln(r/R_0)  \;\;\; ; \;\;\;  \gamma_+(t_+,r) = \gamma_0+\kappa^2 \ln(r/R_0)
  \end{equation}
   \begin{equation}
  \label{equ03a}
   \psi_-(t_-,r) = 0 \;\;\; ; \;\;\; \gamma_-(t_-,r) = 0
 \end{equation}
\begin{equation}
  \label{equ05a}
   R(\tau) = R_0
 \end{equation}

This form can always be achieved with some appropriate choices of additive constants in $\psi_{\pm}$, plus the regularity conditions for $r=0$. With these choices, the matching conditions and evolution equations imply,
\begin{equation}\label{equ06a}
    2 \lambda R_0 -\kappa e^{-\gamma_0} \sqrt{R_0^2+J^2} =0,
\end{equation}

\begin{equation}\label{equ07a}
   R_0 e^{-\gamma_0} -R_0  + 4 \lambda \sqrt{R_0^2+J^2} =0
\end{equation}
and,
\begin{equation}\label{equ08a}
  \lambda R_0^3
- e^{-\gamma_0} J^2 \sqrt{R_0^2+J^2}=0
\end{equation}
which contain also some of the same parameters that appear in (\ref{equ06} - \ref{equ08}). Therefore, these relations impose conditions and constraints on both the initial and final parameters characterizing the corresponding data. Directly from (\ref{equ06a} - \ref{equ08a}) we find,
\begin{eqnarray}
\label{equ10}
  \lambda &=& \frac{R_0 J^2 \sqrt{R_0^2+J^2}}{(R_0^2+2J^2)^2} \nonumber \\
  \kappa &=& 2\frac{J^2}{R_0^2} \\
   e^{-\gamma_0} &=& \frac{R_0^4}{(R_0^2+2J^2)^2}  \nonumber
\end{eqnarray}
and we notice that these imply a further restriction on the range of $\lambda$, namely, for any choice of $R_0$ and $J$ we have $0 \leq \lambda < 0.15879...$. This, however, is not a physical restriction on the data, but rather a consequence of our choice of the value of $\psi$ on $\Sigma$ for the final state. But this choice imposes restrictions on $\psi_i$. To analyze these we introduce $\xi_0$, such that,
\begin{equation}\label{equ11}
    \xi_0 = R_i -R_0
\end{equation}
so that $\xi_0$ gives a measure (and orientation) of the departure of the MSRF configuration from the static configuration. We notice now that (\ref{equ08}) may be written in the form,
\begin{equation}\label{equ12}
\left. \frac{d^2 R}{d \tau^2} \right|_{\tau=0} = -\frac{e^{4 \psi_i} e^{-3 \gamma_i} J^6}{R_0^6 \lambda^2 ( R_0+\xi_0)}
\left(1-\frac{R_0^2e^{2\psi_i}}{(R_0+\xi_0)^2}\right)
\end{equation}
Then the sign of the initial acceleration of the shell will depend on the sign of the term in parenthesis on the right of (\ref{equ12}). To find the relation between this sign and that of $\xi_0$ we may use (\ref{equ11}), (\ref{equ06}) and (\ref{equ07}) to find,
\begin{equation}\label{equ13}
x = \frac{\sqrt{R_0^2+J^2}(R_0^4+4J^4q^2+4R_0^2J^2)}{q(R_0^2+2 J^2)^2\sqrt{R_0^2+q J^2}}
\end{equation}
where,
\begin{equation}\label{equ14}
    x=\frac{R_0+\xi_0}{R_0} \;\;\; ; \;\;\; q = \frac{R_0 e^{\psi_i}}{R_0+\xi_0}
\end{equation}

Now it is easy to check that the right side of (\ref{equ13}) is a monotonically decreasing function of $q$ for $q > 0$, that diverges as $q \to 0^+$, is equal to one for $q=1$ and decreases to,
\begin{equation}\label{equ15}
    \frac{4 J^3\sqrt{R_0^2+J^2}}{(R_0^2+2 J^2)^2} < 1
\end{equation}
for $q \to +\infty$. But this means that $x < 1$ (and, therefore, $\xi_0 < 0$ ) for $q > 1$, when the acceleration is positive, while $x > 1$ (and $\xi_0 > 0$) for $q < 1$, when the acceleration is negative. Thus, we conclude that the initial acceleration is always directed towards the corresponding static radius. Notice that for $q=1$ we have $x=1$, implying $\xi_0=0$ and, therefore, $\psi_i=0$, that is, the static configuration.

We consider now the function $\gamma_+(t_+,r)$. In particular, we want to compare its value at some fixed $r$ for the MSRF data, with that of the assumed final static configuration at the same $r$, since both depend on $r$ only through the term $\ln(r)$, and we have,
\begin{eqnarray}
\label{equ20}
  \exp(\gamma_+(0,r)-\gamma_+(t,r)|_{t_+ \to +\infty}) &=&  \exp(\gamma_i +\kappa^2 \ln(r/R_i)-\gamma_0-\kappa^2 \ln(r/R_0)) \nonumber \\
   &=& \left(\frac{R_0}{R_i}\right)^{\kappa^2} \frac{e^{\gamma_i}}{e^{\gamma_0}} \nonumber \\
   & = &  \left(\frac{R_0}{R_i}\right)^{4 J^2/R_0^4}\frac{R_i q \sqrt{R_0^2+q^2 J^2}}{R_0\sqrt{R_0^2+J^2}} \nonumber
\end{eqnarray}
and, finally,
\begin{equation}\label{equ2}
  \exp(\gamma_+(0,r)-\gamma_+(t_+,r)|_{t_+ \to +\infty})=   \left(\frac{R_0}{R_i}\right)^{4 J^2/R_0^4}\frac{ R_0^4+4 R_0^2 J^2 +4 q^2 J^4}{(R_0^2+2 J^2)^2}
\end{equation}
where $q$ is the same as in Eq. (\ref{equ14}). But, from the previous results, for $\xi_0 < 0$ we have $R_i<R_0$, and $q>1$, while for $\xi_0 > 0$ we have $R_i>R_0$, and $q<1$. Therefore, for $R_i < R_0$ we must have,
\begin{equation}
\label{equ22a}
  \gamma_+(0,r)  >  \gamma_+(t,r)|_{t_+ \to +\infty}
\end{equation}
while for $R_i > R_0$ we must have,
\begin{equation}
\label{equ22b}
  \gamma_+(0,r)  <  \gamma(t_+,r)|_{t_+ \to +\infty}
\end{equation}

This last result, that the function $\gamma_+$ must increase for $R_i > R_0$ to reach the final static configuration, was considered in \cite{nakao} to be in contradiction with the properties of the evolution equations for $\gamma$, and therefore, was considered to imply that in that case the system could not reach a static configuration, implying some sort of instability, possibly related to the dynamic modes of the gravitational field. However, as we show in the Appendix, (\ref{equ22b}) is perfectly compatible with the field equations, and there is no {\em a priori} contradiction with the presumed existence of a static final state. \\

In the next Section we analyze the case where the shell position for the MSRF data is close to static configuration for the same shell. In this case we may consider a linearized expansion about the static configuration and apply some recently obtained results on the dynamics of the Apostolatos -Thorne model \cite{gleramqnm}.

\section{Linearized approximation}

We consider again the MSRF data, assume that we have chosen the appropriate constants so that we have $t_-=t_+=0$ for $\tau=0$, and write it the form,
\begin{eqnarray}
\label{lapp01}
 \left.R \right|_{\tau=0} &=&  R_0 +\epsilon \xi_0 \;\; ;  \;\; \left. \frac{dR}{d\tau} \right|_{\tau=0} = 0 \nonumber \\
\psi_-(0,r) &=& \epsilon \psi_i \;\; ;  \;\;  \gamma_-(0,r)=0 \nonumber \\
 \left. \frac{\partial\psi_-(t_-,r)}{\partial t_-} \right|_{t_-=0} &=& 0 \;\; ;  \;\;  \left.\frac{\partial \gamma_-(t_-,r)}{\partial t_-} \right|_{t_-=0} = 0 \\
 \psi_+(0,r)   &=& \epsilon \psi_i -\kappa\ln \left( \frac{r}{R_0+\epsilon \xi_0}\right)  \;\; ;  \;\;  \gamma_+(0,r)= \gamma_0 +\epsilon \gamma_i+\kappa^2\ln \left( \frac{r}{R_0+\epsilon \xi_0}\right)  \nonumber \\
  \left.\frac{\partial\psi_+(t_+,r)}{\partial t_+} \right|_{t_+=0} &=& 0 \;\; ;  \;\; \left. \frac{\partial \gamma_+(t_+,r)}{\partial t_+} \right|_{t_+=0}= 0 \nonumber
\end{eqnarray}
where $\xi_0$, $\psi_i$, and $\gamma_i$ are some constants, and we have included the auxiliary parameter $\epsilon$ to indicate which quantities are of first order. The set (\ref{lapp01}) corresponds to an exact MSRF data set. We are interested in the case where we have a small departure from the static configuration of the shell, namely, in the limit $\epsilon \to 0$. We may therefore expand (\ref{lapp01}) to first order in $\epsilon$, and consider the resulting set as initial data for the linearized equations of motion for the shell. These were obtained in \cite{gleramqnm}. Briefly stated, one first writes the dynamical variables $R$, $\psi_-$ and $\psi_+$ in the form,

\begin{eqnarray}
\label{gleq01h}
  R(\tau) &=& R_0 + \epsilon \; \xi(\tau) \nonumber \\
  \psi_-(t_-,r) &=& \epsilon\; \chi_1(t_-,r) \\
  \psi_+(t_+,r) &=& - \kappa \ln (r/R_0) + \epsilon \;\chi_2(t_+,r) \nonumber
\end{eqnarray}
where the parameter $\epsilon$ defines the order of the terms, so that the static solution is recovered for $\epsilon=0$. The  $\chi_i$ satisfy the equations,
\begin{eqnarray}
\label{gleq01ha}
 \frac{\partial^2 \chi_1}{\partial t_-^2} -\frac{\partial^2 \chi_1}{\partial r^2}-\frac{1}{r}\frac{\partial \chi_1}{\partial r} &=& 0  \\
 \label{gleq01hb}
 \frac{\partial^2 \chi_2}{\partial t_+^2} -\frac{\partial^2 \chi_2}{\partial r^2}-\frac{1}{r}\frac{\partial \chi_2}{\partial r} &=& 0
\end{eqnarray}
Then, to first order in $\epsilon$ one has,
\begin{eqnarray}
\label{gleq02h}
  \gamma_-(t_-,r) &=& {\cal{O}}(\epsilon^2)   \\
  \gamma_+(t_+,r) &=& \gamma_0 +\kappa^2 \ln (r/R_0) - 2 \epsilon\;\kappa \;\chi_2(t_+,r) \nonumber
\end{eqnarray}
where $\kappa$ and $\gamma_0$ satisfy (\ref{equ10}), and, one finds that it is consistent to this order to set,
\begin{eqnarray}
\label{gleq04h}
 t_-(\tau) &=& \tau + {\cal{O}}(\epsilon)    \\
 t_+(\tau) &=& e^{-\gamma_0} \tau + {\cal{O}}(\epsilon) \nonumber
\end{eqnarray}
In fact, only the zeroth order terms in $t_{\pm}$ appear in the linearized equations, and, instead of (\ref{gleq01ha}), one has,
\begin{eqnarray}
\label{gleq01ha1}
 \frac{\partial^2 \chi_1}{\partial \tau^2} -\frac{\partial^2 \chi_1}{\partial r^2}-\frac{1}{r}\frac{\partial \chi_1}{\partial r} &=& 0  \\
 \label{gleq01hb1}
 \frac{(2 J^2+R_0^2)^4}{R_0^8} \frac{\partial^2 \chi_3(\tau,r)}{\partial \tau^2}-\frac{\partial^2 \chi_3(\tau,r)}{\partial r^2}-\frac{1}{r}\frac{\partial \chi_3(\tau,r)}{\partial r}  &=& 0
\end{eqnarray}
where,
\begin{equation}\label{gleq07h}
\chi_3(\tau,r)=\chi_2(e^{-\gamma_0}\tau,r)
\end{equation}
These equations describe the dynamics of the radiative part of the gravitational field. The corresponding linearized equations for the motion of the shell are, (see \cite{gleramqnm} for details),
\begin{eqnarray}
\label{eq05h}
0 & = & {\frac {2 {J}^{4} \left( 6\,{R_{{0}}}^{2}{J}^{2}+4{J}^{4}+3{R_{{0
}}}^{4} \right)   }{ \left( {R_{{0}}}^{2}+{J}^{2
} \right)  \left( 2{J}^{2}+{R_{{0}}}^{2} \right) ^{2}{R_{{0}}}^{4}}} \xi \left( \tau \right)  \nonumber \\
& &
-{\frac {2 R_{{0}}{J}^{4}  }
{ \left( {R_{{0}}}^{2}+{J}^{2} \right)  \left( 2{J}^{2}+{R_{{0}}}^{2
} \right) ^{2}}} \chi_1 \left( \tau,R_{{0}} \right)
- \left.{\frac {\partial \chi_1 \left(\tau,r \right)}{\partial r}}\right|_{r=R_0}
 \\
 & &
-{\frac {2{J}^{2} \left( 4{J}^{2}+{R_{{0}}}^{2}
 \right)   }{ \left( 2{J}^{
2}+{R_{{0}}}^{2} \right) ^{2}R_{{0}}}}\chi_3 \left(\tau,R_{{0}} \right)
+{\frac {   {R_{{0}}}^{4}}{ \left( 2\,{J}^{2}+{R_{{0}}}^{2}
 \right) ^{2}}}
 \left.{\frac {\partial \chi_3 \left(\tau,r \right)}{\partial r}}\right|_{r=R_0}, \nonumber \\
0 & = & {\frac {2 {J}^{2}  }{{R_{{0}}}^{3}}}\xi \left( \tau \right)
+\chi_1 \left( \tau,R_{{0}} \right)
-\chi_3 \left(\tau,R_{{0}}
 \right) \nonumber
\end{eqnarray}
corresponding to the matching conditions, and an equation of motion for $\xi(\tau)$,
\begin{eqnarray}
\label{eq06h}
{\frac {d^{2}}{d{\tau}^{2}}}\xi \left( \tau
 \right) & = & -{\frac {{J}^{2} \left( 4{J}^{6}+6{R_{{0}}}^{2}{J}^{4}+5{J}^{2}{R
_{{0}}}^{4}+2{R_{{0}}}^{6} \right)  }{
 \left( 2\,{J}^{2}+{R_{{0}}}^{2} \right) ^{2} \left( {R_{{0}}}^{2}+{J}
^{2} \right) ^{2}{R_{{0}}}^{2}}}  \xi \left( \tau \right) \nonumber\\
& & +{\frac {{J}^{2}{R_{{0}}}^{3} \left( 3
{J}^{2}+2{R_{{0}}}^{2} \right)   }{ \left( 2{J}^{2}+{R_{{0}}}^{2} \right) ^{2} \left( {R_{{0
}}}^{2}+{J}^{2} \right) ^{2}}} \chi_1 \left( \tau,R_{{0}}
 \right)
 +{\frac {   {R_{{0}}}^{2}}{{R_
{{0}}}^{2}+{J}^{2}}}
\left.{\frac {\partial \chi_{{1}} \left( \tau,r \right)}{\partial r }}
 \right|_{r=R_0}  \\
 & &
+{\frac { \left( 4{J}^{2}+{R_{{0}}}^{2
} \right) R_{{0}}{J}^{2}  }{
 \left( {R_{{0}}}^{2}+{J}^{2} \right)  \left( 2{J}^{2}+{R_{{0}}}^{2}
 \right) ^{2}}}\chi_3 \left(\tau,R_{{0}} \right) \nonumber
\end{eqnarray}

It was shown in \cite{gleramqnm} that associated with the solutions of the system of linearized equations there exists a positive definite constant of the motion given by,
\begin{eqnarray}
\label{stab07}
E_s & = &  \frac{1}{2}\left(\frac {d \xi}{d\tau}\right)^2
+{\frac {{J}^{2}  }{6\,{R_{{0}}}^{2}{J}^{2}+4\,{J}^{4}+{R_{{0}}}^{4}}} \xi^2\nonumber \\
& &+{\frac {{R_{{0}}}^{14}   }{8 \left( 2{J}^{2}+{R_{{0}}}
^{2} \right) ^{2}{J}^{4} \left( 6{R_{{0}}}^{2}{J}^{2}+4{J}^{4}+{R_
{{0}}}^{4} \right) }}
\left. \left( \dfrac{ \left( 2{J}^{2}+{R_{{0}}}^{2} \right) ^{2
}}{R_0^4}{\dfrac {\partial \chi_{{1}}}{\partial r}}  -
   {\dfrac {\partial \chi_{{3}}}{\partial r}}\right)^2 \right|_{r=R_0} \nonumber
 \\
& & +{\frac {{R_{{0}}}^{4}  }{ 2\left( {R_{{0}}}^{2}+{J}^{2} \right) {J}^{2}}}
 \int_0^{R_0}\frac{r}{2}\left[\left(\frac{\partial \chi_1}{\partial \tau}\right)^2 +\left(\frac{\partial \chi_1}{\partial r}\right)^2 \right] dr \\
& & +{\frac {{R_{{0}}}^{8}   }{ 2\left( {R_{{0}}}^{2}+{J}^{
2} \right)  \left( 2\,{J}^{2}+{R_{{0}}}^{2} \right) ^{2}{J}^{2}}}
 \int_{R_0}^{\infty}\frac{r}{2}\left[\frac{(2 J^2+R_0^2)^4}{R_0^8}\left(\frac{\partial \chi_3}{\partial \tau}\right)^2 +\left(\frac{\partial \chi_3}{\partial r}\right)^2 \right] dr  \nonumber
\end{eqnarray}

To apply these results to our problem we expand (\ref{lapp01}) to first order in $\epsilon$ to obtain the corresponding initial data for $\xi(\tau)$, $\chi_1(\tau,r)$, and $\chi_3(\tau,r)$. The results are,
\begin{eqnarray}
\label{gl30}
  \xi(0) &=& \xi_0 \;\;  ;  \;\; \left.\frac{d \xi}{d \tau} \right|_{\tau=0} = 0 \nonumber \\
  \chi_1(0,r) &=& \chi_i \;\;  ;  \;\; \left.\frac{\partial \chi_1}{\partial \tau} \right|_{\tau=0} = 0 \\
  \chi_3(0,r) &=& \chi_i +\frac{2 J^2}{R_0^3} \xi_0 \;\;  ;  \;\; \left.\frac{\partial \chi_3}{\partial \tau} \right|_{\tau=0} = 0 \nonumber
\end{eqnarray}
where $\xi_0$ and $\chi_i$ are not independent, but we have the relation,
\begin{equation}\label{gl31}
    \chi_i = \frac{J^2 (R_0^4-4J^4 -4J^2 R_0^2)}{R_0^3(6 R_0^2J^2+4J^4+R_0^4)} \xi_0
\end{equation}

If we apply now this data to compute $E_s$ at $\tau=0$ we find,
\begin{equation}
\label{gl32}
E_s =
{\frac {{J}^{2}  }{6\,{R_{{0}}}^{2}{J}^{2}+4\,{J}^{4}+{R_{{0}}}^{4}}} \xi_0^2
\end{equation}
and therefore, we must have $\xi(\tau)^2 \leq \xi_0$ for all $\tau$, and the motion of the shell is always bounded, proving the (linear) stability of the shell under MSRF initial data. Moreover, since for $\tau>0$ the region $r>R_0$ will contain outgoing radiation, we expect a steady decrease in the maximum amplitude for $\xi$, and an eventual approach to $\xi =0 $, corresponding to the static solution.

The existence of $E_s$ does not provide direct information on the rate of approach to $\xi =0 $. To obtain this information we may proceed as follows. We first notice that after some initial transitory motion, possibly dominated by  quasi normal ringing oscillations, these oscillations would damp out since all available energy for motion of the shell will eventually be radiated to infinity, as the system approaches its final static configuration. In this situation the field outside the shell would approach the situation described in the Appendix, for the late time behaviour of any solution that approaches asymptotically a constant, (Eq. (\ref{ap15a})), namely, for $\tau >> r$, in particular for $\tau \to \infty$, and for $r$ of the order of $ R_0$, we should have,
\begin{equation}
\label{gl50a}
\chi_2(t_+,r) \sim {\cal{A}} \frac{\ln(r)}{\ln(t_+)},
\end{equation}
where ${\cal{A}}$ is of the order of (the constant value) $\chi_3(0,r)$. More generally, this implies that for sufficiently large $\tau$, and $r << \tau$,
\begin{equation}
\label{gl50}
 \chi_3(\tau,r) \sim   \frac{A \ln(r/R_0) +B}{\ln(e^{-\gamma_0}\tau)+Q}
\end{equation}
where $A$, $B$ and $Q$ are constants, is an asymptotic approximation to the solution of our problem that approaches (\ref{gl50a}) for $\tau \to \infty$. Before replacing (\ref{gl50}) in the equations of motion we may use (\ref{eq05h}) to write $\chi_1$ and $\partial \chi_1/\partial r$ in terms of $\xi$, $\chi_3$ and $\partial \chi_3/\partial r$, and rewrite (\ref{eq06h}) in the alternative form,
 \begin{eqnarray}
 \label{gl60}
  {\frac {d^{2}\xi}{d{\tau}^{2}}} & = & {\frac { \left( 2{J}^{4}+{{R_0}}^{2}{J}^{2}-2{{R_0}}^{4} \right) {J}
^{2} }{ \left( 2{J}^{2}+{{R_0}}^{2}
 \right)  \left( {{R_0}}^{2}+{J}^{2} \right) ^{2}{{R_0}}^{2}}} \xi
 -{
\frac {{J}^{2}{R_0}\, \left( 4{{R_0}}^{2}{J}^{2}+4{J}^{4}-{{
R_0}}^{4} \right)     }{ \left( 2
{J}^{2}+{{R_0}}^{2} \right) ^{2} \left( {{R_0}}^{2}+{J}^{2}
 \right) ^{2}}} \chi_{{3}} \nonumber \\
 & & +{\frac {   {{R_0}}^{6}}{ \left( {{R_0}}^{2}+{J}^{2} \right)  \left( 2\,{J}^{2}+{{R0_0}}^{2} \right) ^{2
}}} {\frac {\partial \chi_3}{\partial r}}
 \end{eqnarray}

Then, for large $\tau$ we have,\cite{nota1}
\begin{equation}\label{gl70}
\xi(\tau) \sim-{\frac {{R_0}^{3} \left( {R_0}^{4} \left( {R_0}^{2}+{J
}^{2} \right) A-B{J}^{2} \left( 4{J}^{2}{R_0}^{2}+4\,{J}^{4}-{
R_0}^{4} \right)  \right) }{{J}^{2} \left( 2{J}^{2}+{R_0}^{
2} \right)  \left( 2{J}^{4}+{J}^{2}{R_0}^{2}-2{R_0}^{4}
 \right)  \left( \ln  \left( {{\rm e}^{-\gamma_{{0}}}}\tau \right) +Q
 \right) }}
\end{equation}

We finally notice that since both $\chi_3$ and $\xi$ vanish in the limit $\tau \to \infty$, we have the same limit for both $\chi_1$ and $\partial \chi_1/\partial r$ for $r=R_0$, and since $\chi_1$ is regular in $0 \leq r \leq R_0$ we must also have $\chi_1(t_-,r) \to 0$, in the limit $\tau \to \infty$, and the system approaches asymptotically the static configuration.

As we shall show in the next Section, these results are supported by a numerical integration of the equations of motion. In fact, for the purpose of comparison with the numerical results, it will be useful to notice that, in general we should have,
\begin{equation}\label{gl80}
    \chi_3(\tau,r) \sim   A_1 \ln(r) +B_1
\end{equation}
where $A_1$ and $B_1$ are constants, for the $r$ dependence of $\chi_3$ for fixed $\tau >> r$, and,
\begin{equation}\label{gl82}
   \frac{1}{ \chi_3(\tau,r)} \sim   A_2 \ln(\tau) +B_2
\end{equation}
where $A_2$ and $B_2$ are constants, for the $\tau$ dependence of $\chi_3$, for $\tau >> r$, and some fixe $r$, for instance $r=R_0$.

Similarly, we should have
\begin{equation}\label{gl84}
   \frac{1}{ \xi(\tau)} \sim   A_3 \ln(\tau) +B_3
\end{equation}
where $A_3$ and $B_3$ are constants, for the $\tau$ dependence of $\xi$, for sufficiently large $\tau$. Notice that $A_1$, and $B_1$, as defined, depend on $\tau$. In any case, all the quantities $A_i$, and $B_i$ are related by Eqs. (\ref{gl50}) and (\ref{gl70}), but we shall not make use of the explicit expressions here.

\section{Numerical examples}

The previous analysis does not give detailed information on the form in which the system, and in particular the shell, actually evolves towards a static configuration starting from MSRF initial data. It is then interesting to study the initial part of this process, in particular, to see if, in accordance with the qualitative arguments of Apostolatos and Thorne \cite{apostol}, this approach is dominated by an oscillatory part (quasi-normal ringing) that dampens out as the shell approaches its final static radius. Furthermore, a simple intuition would probably dictate that this oscillations are about the final static radius, effectively approaching this radius with the exponential decay characteristic of the quasi normal ringing. As we shall find in the examples analyzed below, this is partly the case, but there are some unexpected features that appear in the evolution of the system, that are in accord with the analysis of the final approach described in the previous Section.

As indicated in \cite{gleramqnm}, it is easy to set up a numerical procedure to integrate the linearized equations as an initial plus boundary value problem, at least in a finite region of $r$ that includes $r=0$. If the region extends to $r=r_o$, then the integration can be extended to a time $\tau$ of the order of $r_o$, so that we can explore the evolution for a larger time by simply choosing a larger value for $r_o$. We refer to \cite{gleramqnm} for further details.

As a first example we consider the case $R_0=1$, $J=0.5$, and place $r_o = 1400$. The initial displacement is $\xi_0=0.1$. This corresponds to $\chi_3(0,r)=0.0477...$. In Figure 1 we display $\xi(\tau)$ as a function of $\tau$. We notice that there is initially a damped oscillation, not about $\xi=0$, but, rather, about a position that decreases rapidly at the beginning, but eventually tends to approach $\xi=0$ very slowly in time, even after the oscillation has essentially damped out. Next, in Fig. 2, we plot $\xi(\tau)^{-1}$ as a function of $\ln(\tau)$. The linear dependence on $\ln(\tau)$ given by (\ref{gl84}) is clearly seen for the larger values of $\tau$. In Figure 3 we display $\chi_3(\tau,r)$ as a function of $r$, for $\tau=1400$. Here the effect of the initial oscillation of $\xi$ appears as an outgoing wave for $r$ of the order of $600$, in accordance with the fact that, from (\ref{gleq01hb1}), these waves propagate with a speed $v=R_0^4/(R_0^2+J^2)^2 = 0.4444...$. After the wave essentially dampens out we see that $\chi_3(\tau,r)$ decreases monotonically to a value close to zero as $r$ approaches $R_0$. A more detailed view of the ourgoing wave egion is depicted in Fig. 4. In Figure 5 we plot $\chi_3(\tau,r)$ as a function of $\ln(r)$. This shows that in the region of $r$ between $R_0$ and the outgoing wave, and for fixed $\tau$, $\chi_3(\tau,r)$ depends linearly on $\ln(r)$, that is, we have $\chi_3(\tau,r) \sim A + B \ln(r)$ for some constants $A$ and $B$, in agreement with the discussion in the previous Section, and (\ref{gl80}). Finally, in Fig. 6, we display the results of the numerical integration for $1/\chi_3(\tau,R_0)$. The linear dependence on $\ln(\tau$ given by (\ref{gl82}) is evident in this figure.

\begin{figure}
\centerline{\includegraphics[height=12cm,angle=0]{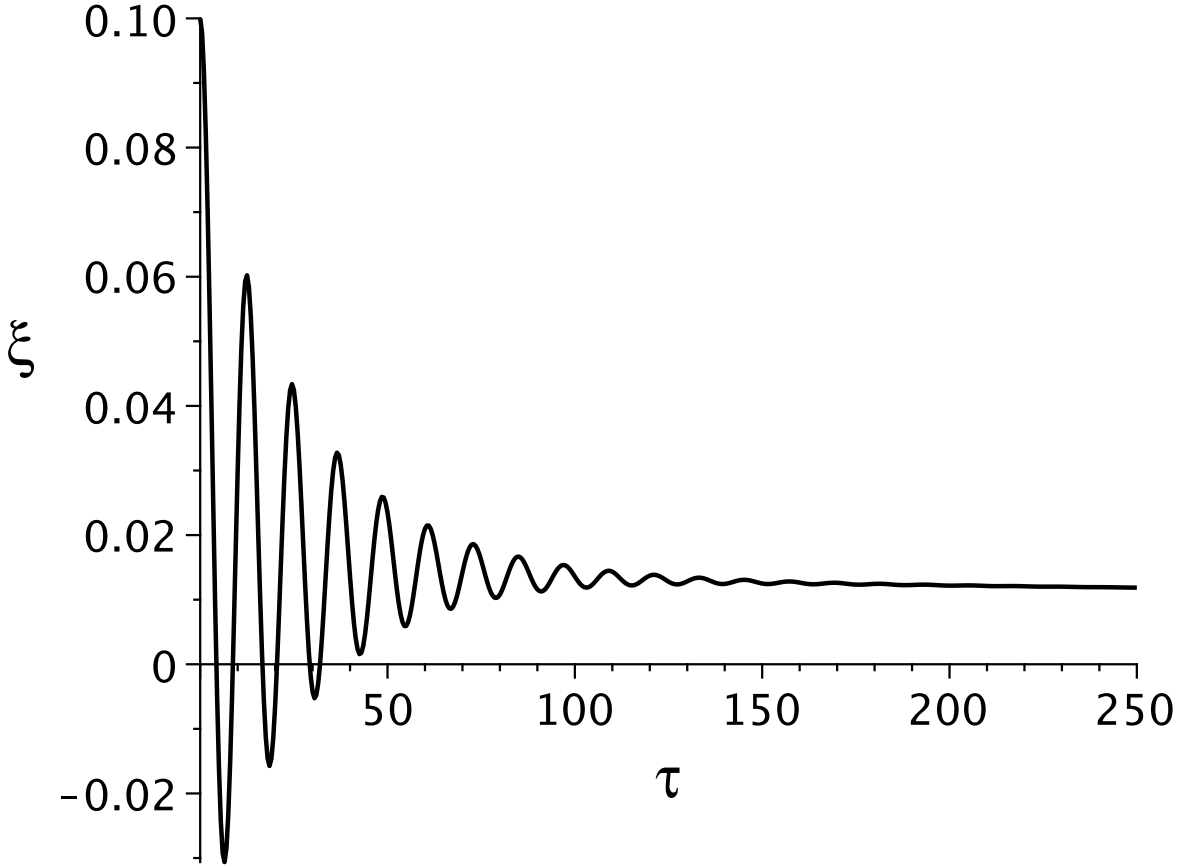}}
\caption{$\xi(\tau)$ as a function of $\tau$, for a shell with $R_0=1.0$, $J=0.5$ and an initial displacement $\xi_0=0.1$. We notice that there is initially a damped oscillation, not about $\xi=0$, but, rather, about a position with $\xi \neq 0$, even after the oscillation has essentially damped out. Here $\tau$ is approximately in the range $0 \leq \tau < 300$.}
\end{figure}

\begin{figure}
\centerline{\includegraphics[height=12cm,angle=0]{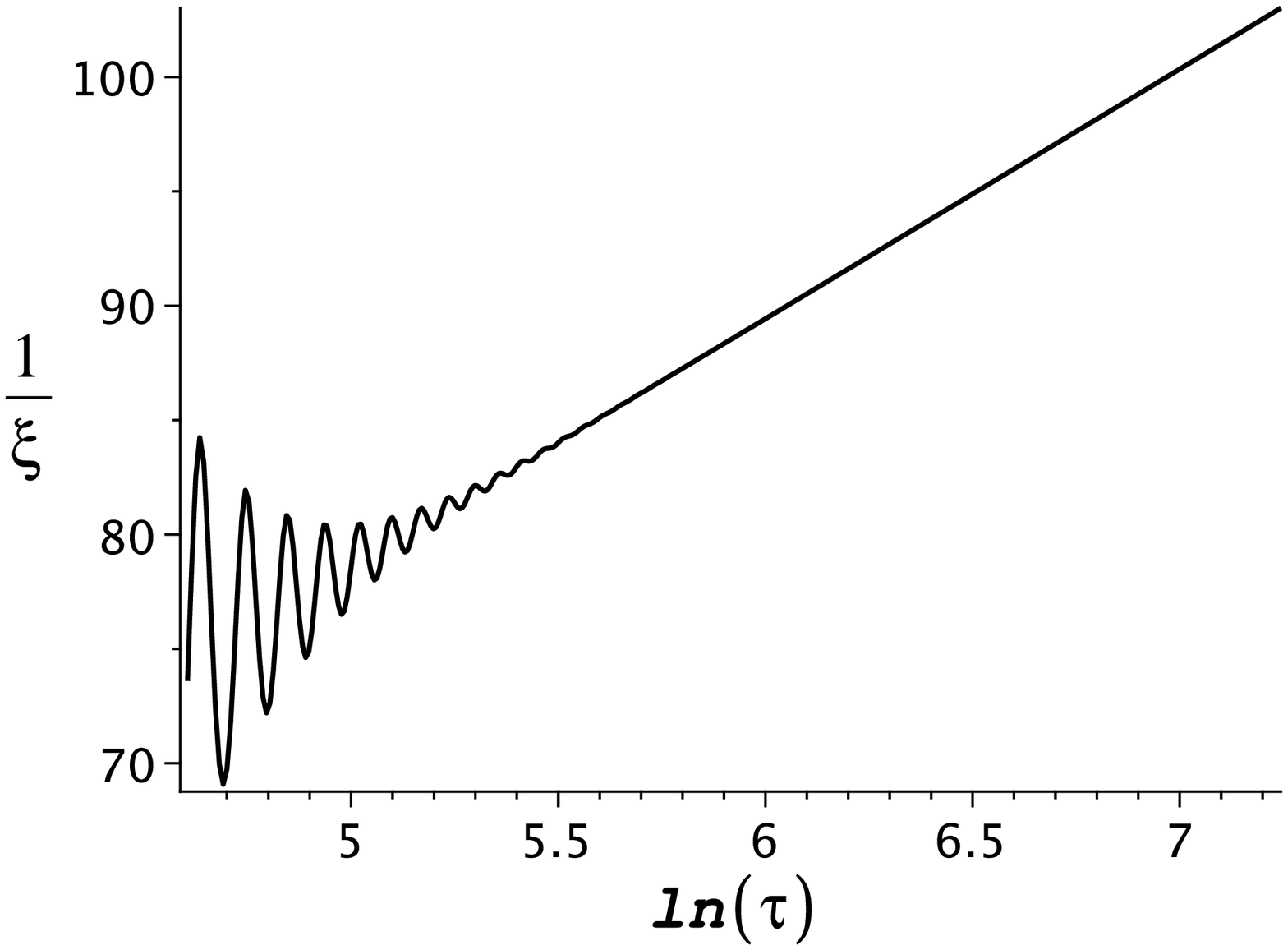}}
\caption{$1/\xi(\tau)$ as a function of $ln(\tau)$, for a shell with $R_0=1.0$, $J=0.5$ and an initial displacement $\xi_0=0.1$.  Here the range of $\tau$ is approximately $200 < \tau < 1500$. The linear dependence of $1/\xi(\tau)$ on $ln(\tau)$, for sufficiently large $\tau$ is evident in the graph.}
\end{figure}

\begin{figure}
\centerline{\includegraphics[height=12cm,angle=0]{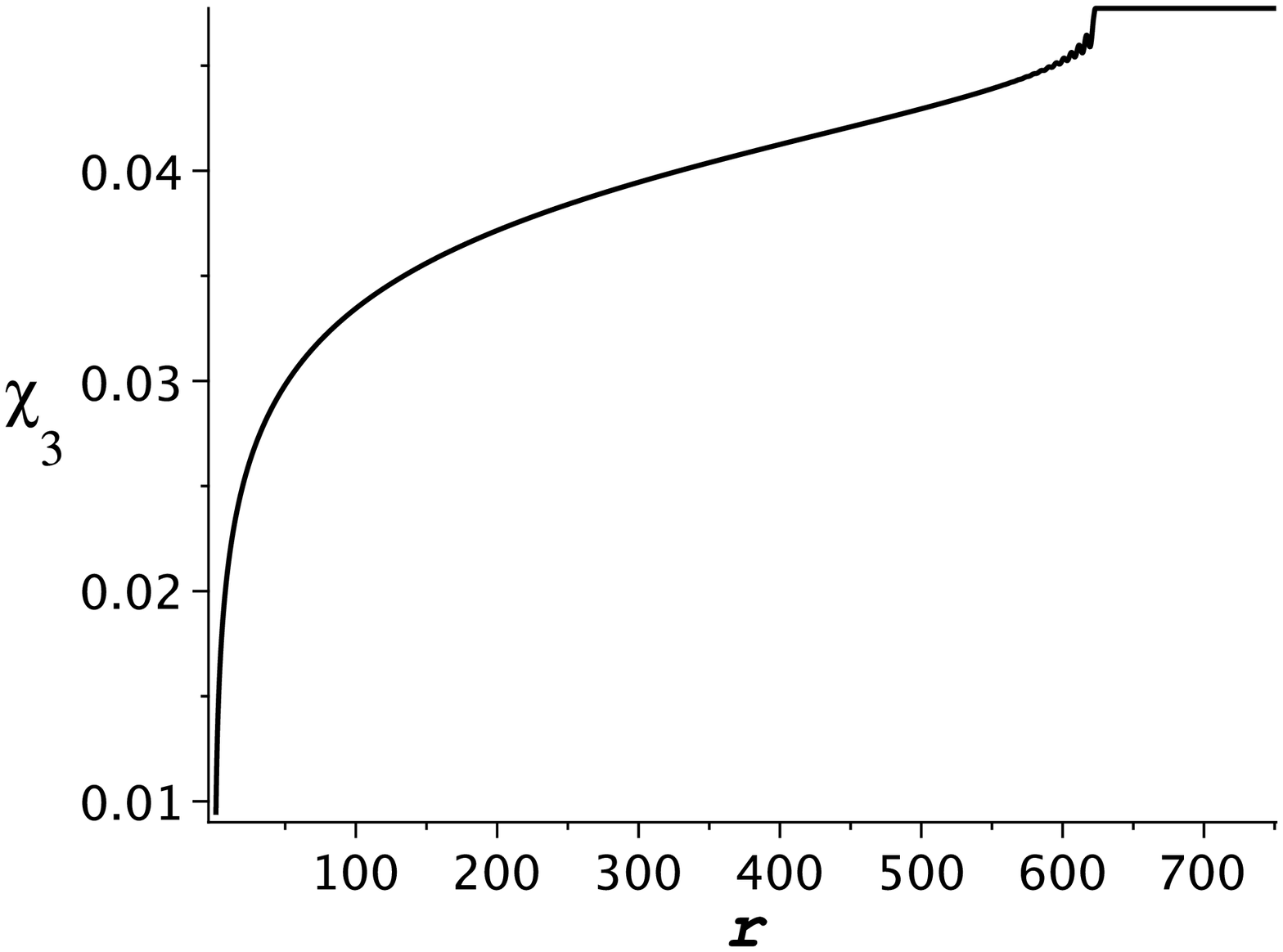}}
\caption{$\chi_3(\tau,r)$ as a function of $r$, for $\tau=1400$, for the same shell and initial conditions as in Figure 1. Here the effect of the initial oscillation of the shell appears as an outgoing wave for $r$ of the order of $620$.}
\end{figure}

\begin{figure}
\centerline{\includegraphics[height=12cm,angle=0]{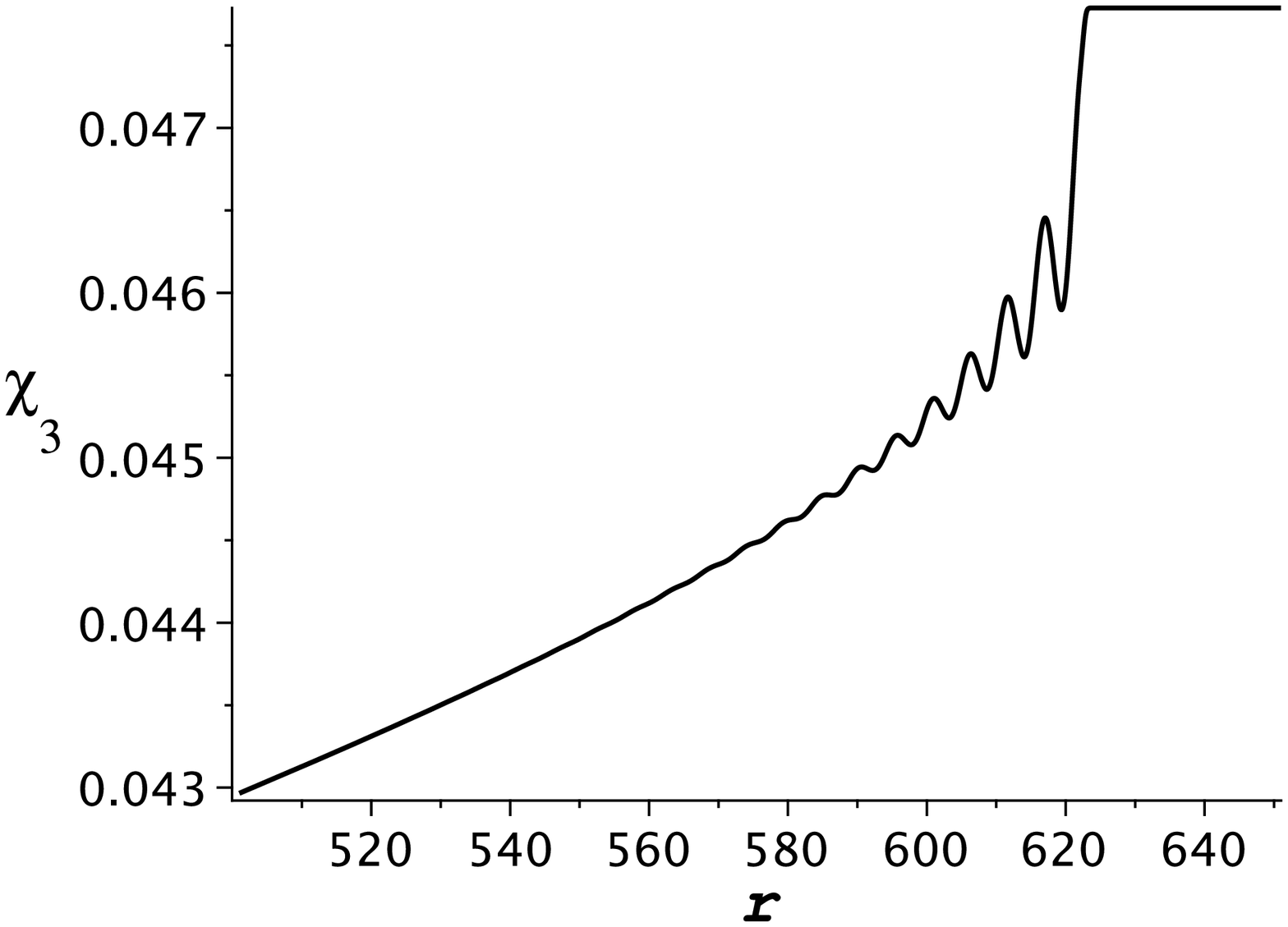}}
\caption{$\chi_3(\tau,r)$ as a function of $r$, for $\tau=1400$, for the same shell and initial conditions as in Figure 1. This is an enlarged view of the region $500 < r < 650$ of Figure 3, to show details of the outgoing wave part of $\chi_3(\tau,r)$.}
\end{figure}

\begin{figure}
\centerline{\includegraphics[height=12cm,angle=0]{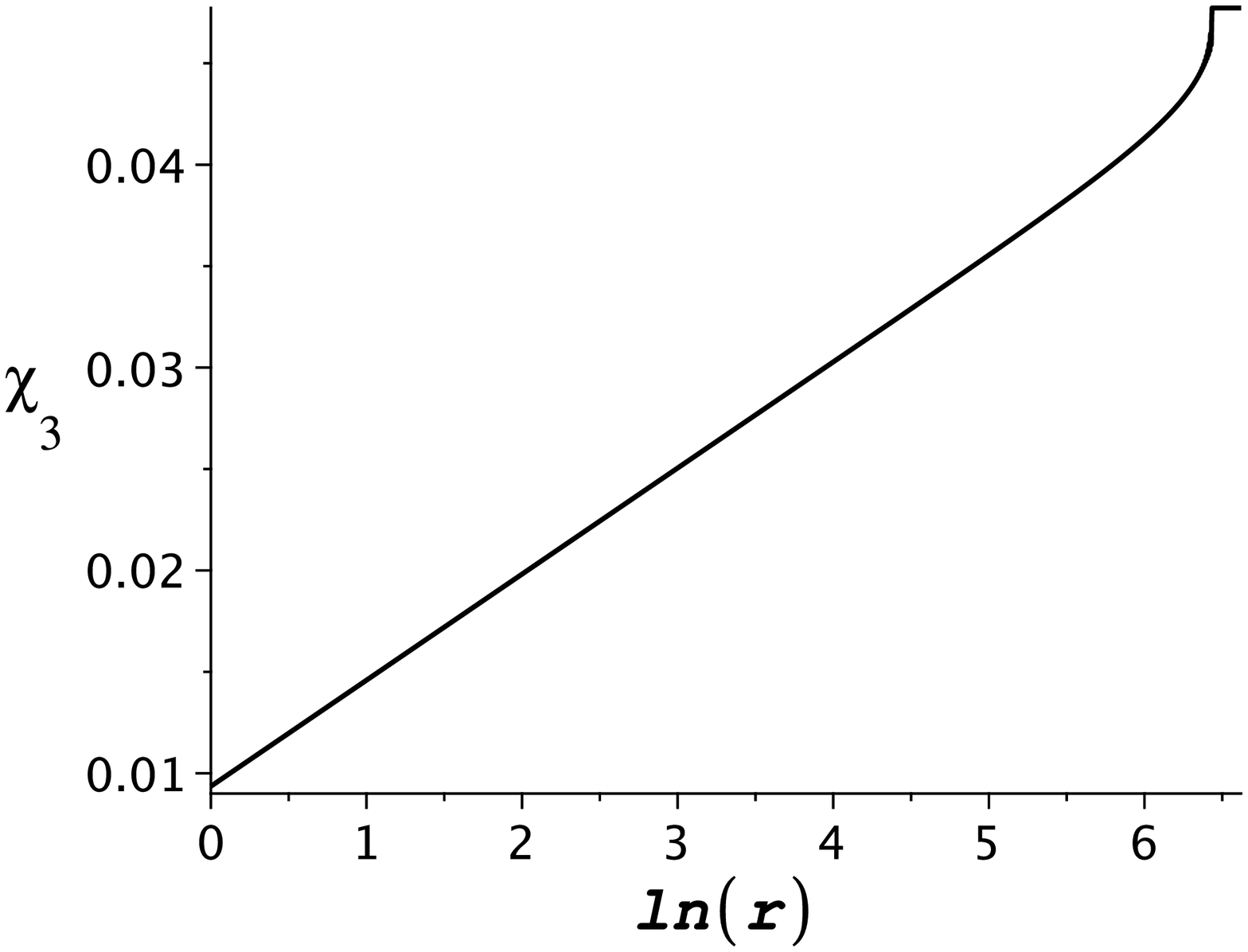}}
\caption{$\chi_3(\tau,r)$ as a function of $\ln(r)$, for $\tau=1400$, for the same shell and initial conditions as in Figure 1. The linear dependence on $\ln(r)$ in the region of $r$ between $R_0$ and the outgoing wave is evident.}
\end{figure}

\begin{figure}
\centerline{\includegraphics[height=12cm,angle=0]{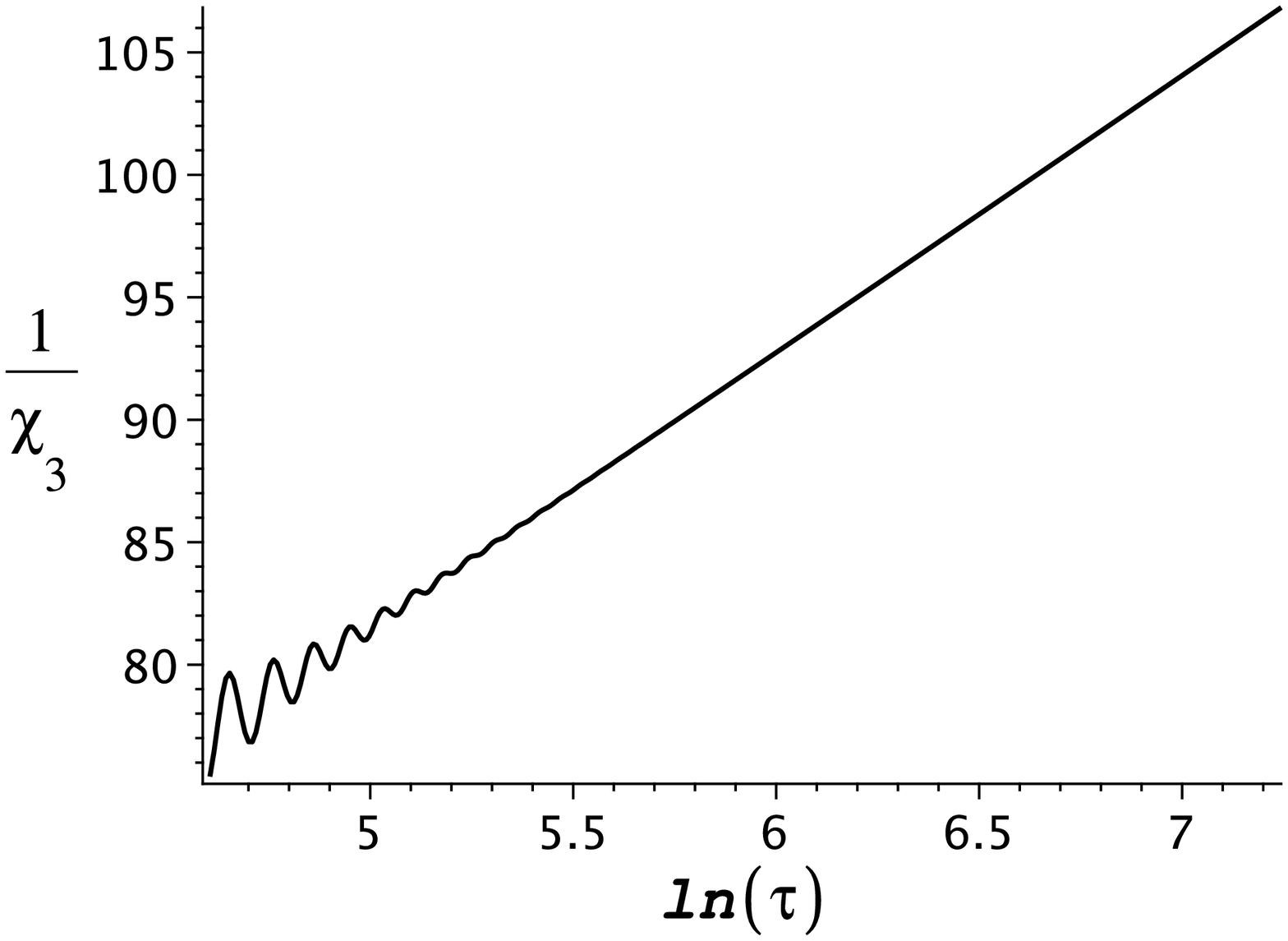}}
\caption{$1/\chi_3(\tau,r)$ as a function of $\ln(\tau)$, for $r=R_0$, for the same shell and initial conditions as in Figure 1. The linear dependence on $\ln(\tau)$ at late times is clearly seen.}
\end{figure}

As a second example we consider a shell with parameters $R_0=1$, and $J=1$. The initial displacement is $\xi_0=-0.1$. This corresponds to $\chi_3(0,r)=-0.136...$. We placed again $r_o=1400$. Figure 7 shows again $\xi(\tau)$ as a function of $\tau$. The behaviour is qualitatively similar to the previous example, but now we have a much stronger damping of the initial oscillations, although, again, we find for $\xi(\tau)$ a slow decrease in absolute value as $\tau$ increases after the oscillations damp out. Notice that now we have set a negative initial value for $\xi$, and that in this case $\xi$ approaches zero from negative values. In Fig. 8 we see the linear dependence of $\xi(\tau)^{-1}$ on $\ln(\tau)$, just as in the case shown in Fig.2, but, of course, with different parameters. In Fig. 9 we show the result of the numerical integration for $\chi_3(\tau,r)$ as a function of $r$, for $\tau=1400$. The qualitative features on this figure are similar to those on Fig.3, although now $\chi_3 <0$ because of our choice of $\xi_0$. We also find the outgoing wave at $r \sim 150$, because the speed of the waves (for $\chi_3(\tau,r)$) is now only $v = 0.111...$. Fig. 10 gives a more detailed view of the wave zone, and Figs. 11 and 12, display respectively the linear dependence of $\chi_3$ on $\ln(r)$ for fixed $\tau >> r$, and for
$1/\chi_3(R_0,tau)$ on $\ln(tau)$, in complete agreement with the discussion in the previous Section, and those obtained numerically for $R_0=$, $J=0.5$. We should also mention that the numerical values of the parameters obtained by simply fitting these linear dependencies are consistent with the relations implied by (\ref{gl50}) and (\ref{gl70}), but we shall not give details here.

\begin{figure}
\centerline{\includegraphics[height=10cm,angle=0]{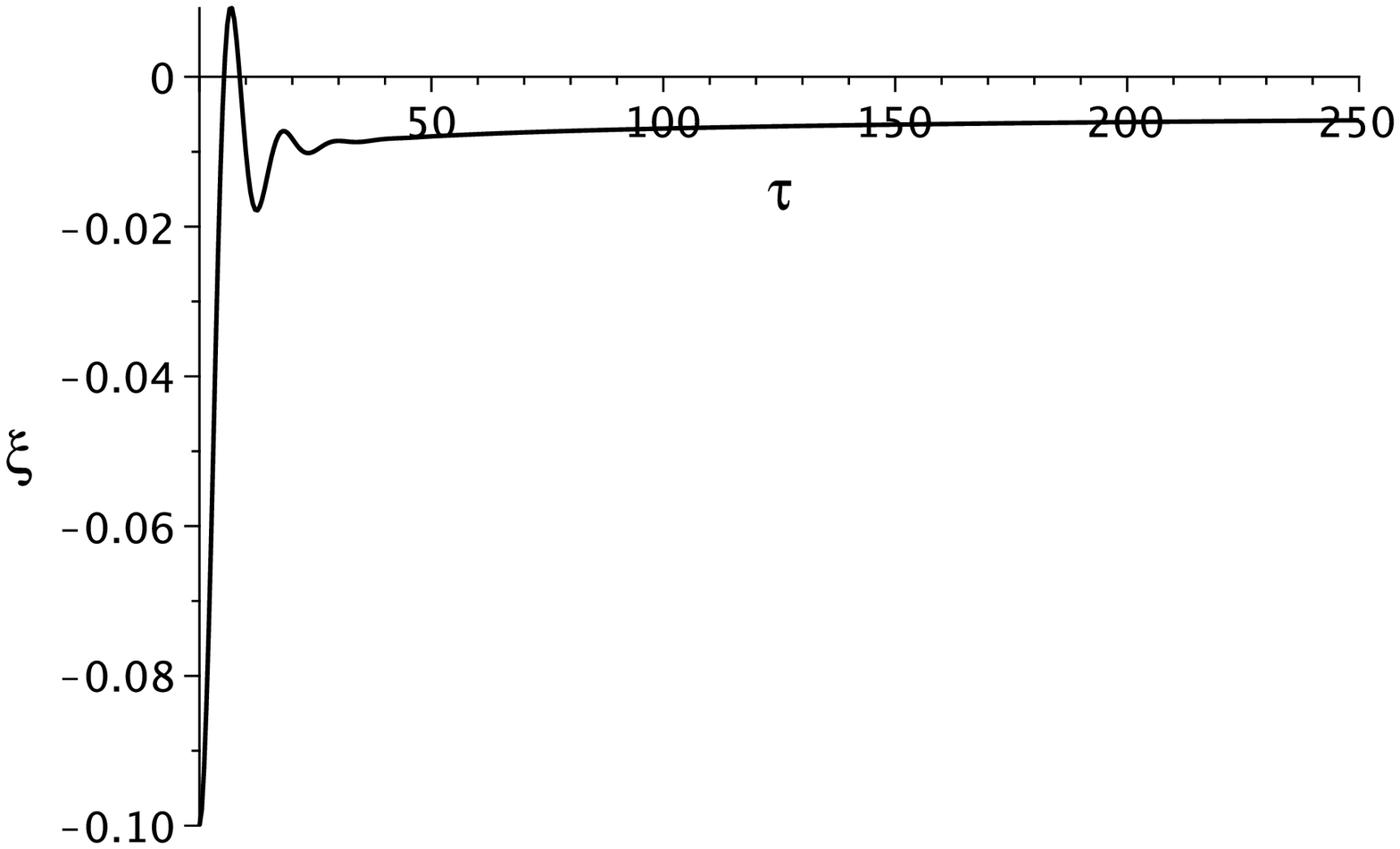}}
\caption{$\xi(\tau)$ as a function of $\tau$, for a shell with $R_0=1.0$, $J=1.0$ and an initial displacement $\xi_0=-0.1$. We notice that initially we have a strongly damped oscillation, followed by a slow decrease in the absolute value of $\xi(\tau)$ . Here $\tau$ is approximately in the range $0 \leq \tau < 250$.}
\end{figure}

\begin{figure}
\centerline{\includegraphics[height=10cm,angle=0]{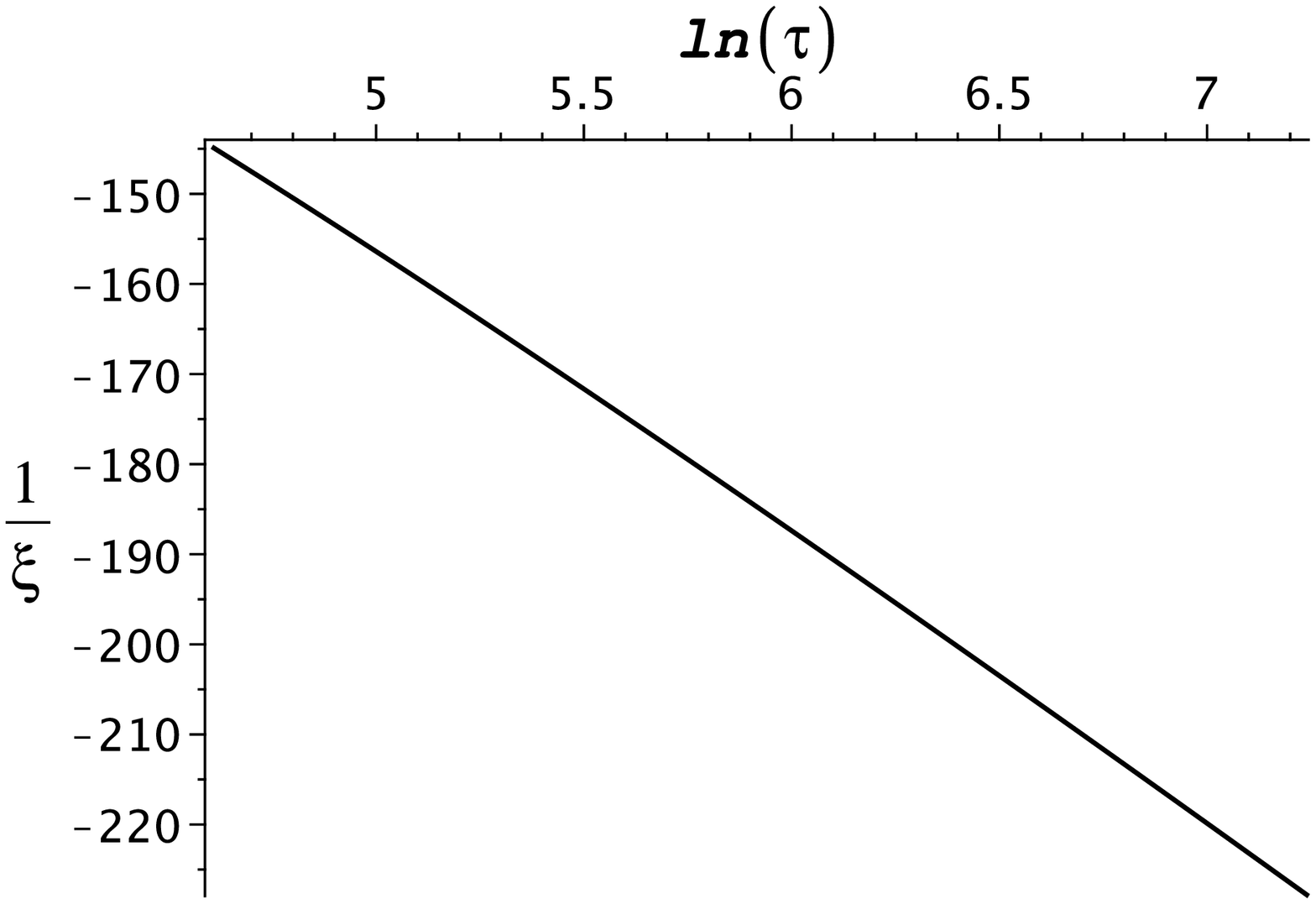}}
\caption{$1/\xi(\tau)$ as a function of $ln(\tau)$, for a shell with $R_0=1.0$, $J=1.0$ and an initial displacement $\xi_0=-0.1$.  Here the range of $\tau$ is approximately $200 < \tau < 1500$. The linear dependence of $1/\xi(\tau)$ on $ln(\tau)$, for sufficiently large $\tau$ is evident in the graph.}
\end{figure}

\begin{figure}
\centerline{\includegraphics[height=10cm,angle=0]{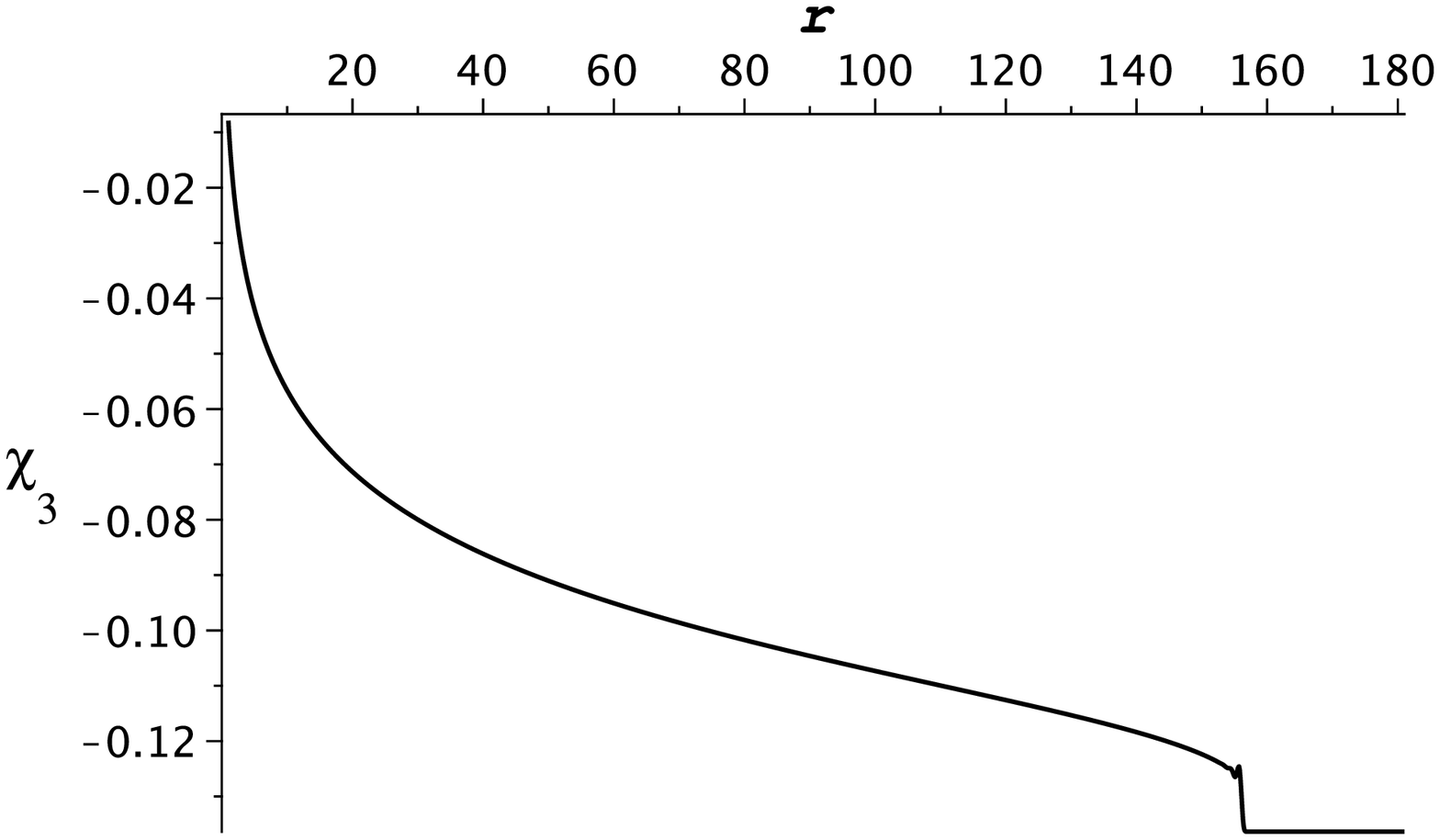}}
\caption{$\chi_3(\tau,r)$ as a function of $r$, for $\tau=1400$, for the same shell and initial conditions as in Figure 7. Here the effect of the initial oscillation of the shell appears as an outgoing wave for $r$ of the order of $150$.}
\end{figure}

\begin{figure}
\centerline{\includegraphics[height=10cm,angle=0]{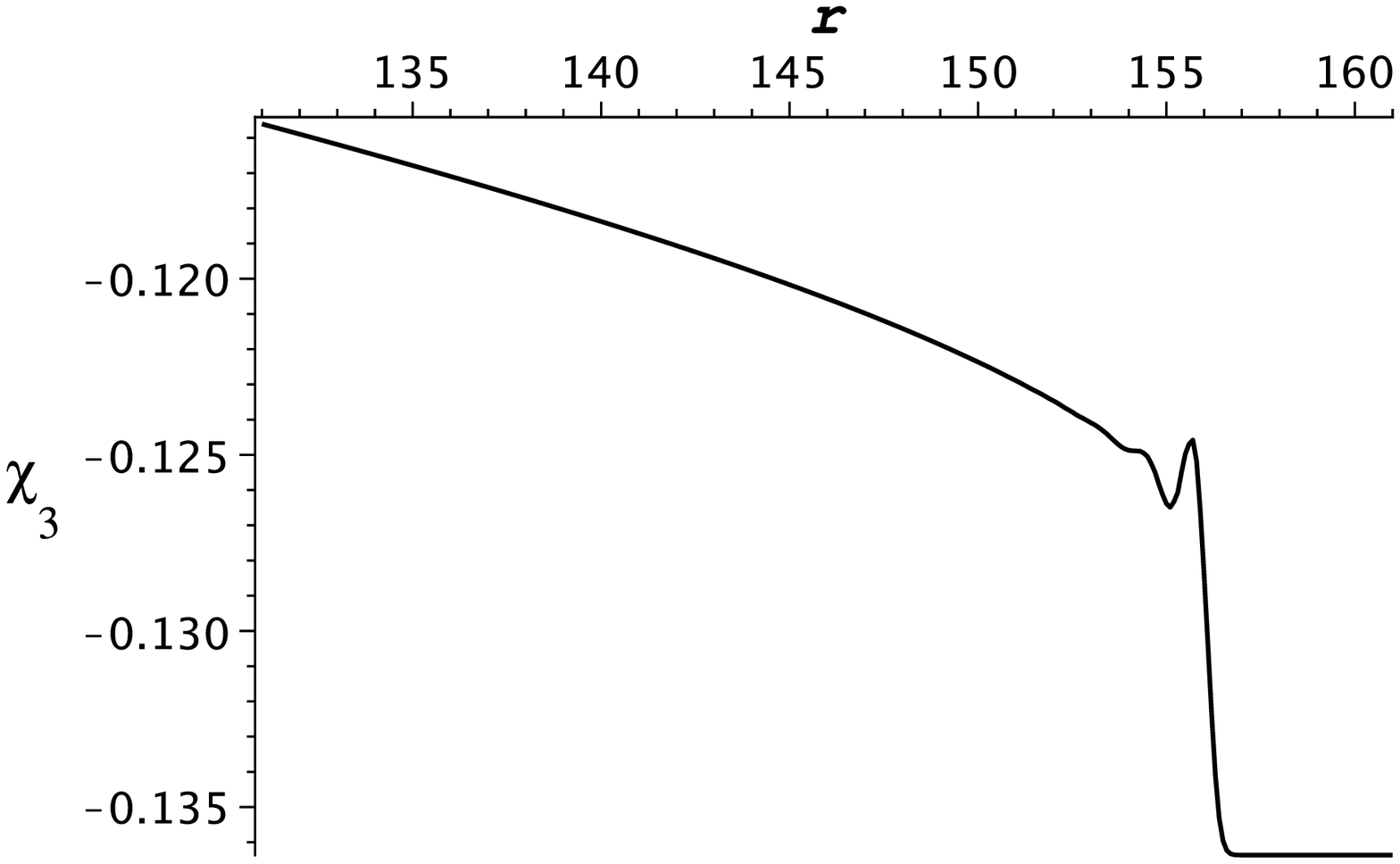}}
\caption{$\chi_3(\tau,r)$ as a function of $r$, for $\tau=1400$, for the same shell and initial conditions as in Figure 7. This is an enlarged view of the region $130 < r < 160$ of Figure 9, to show details of the outgoing wave part of $\chi_3(\tau,r)$.}
\end{figure}

\begin{figure}
\centerline{\includegraphics[height=10cm,angle=0]{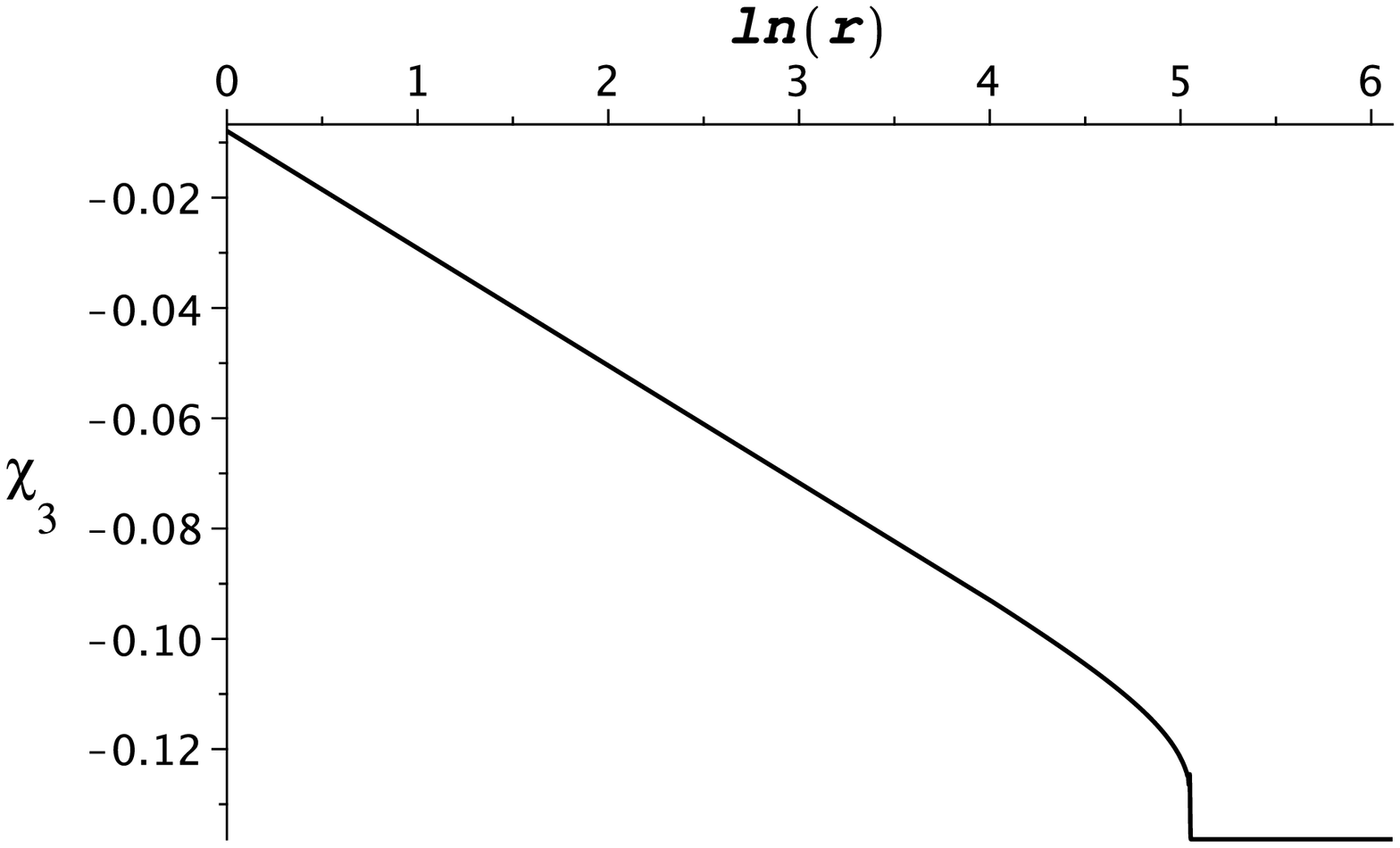}}
\caption{$\chi_3(\tau,r)$ as a function of $\ln(r)$, for $\tau=1400$, for the same shell and initial conditions as in Figure 7. The linear dependence on $\ln(r)$ in the region of $r$ between $R_0$ and the outgoing wave is again evident.}
\end{figure}

\begin{figure}
\centerline{\includegraphics[height=10cm,angle=0]{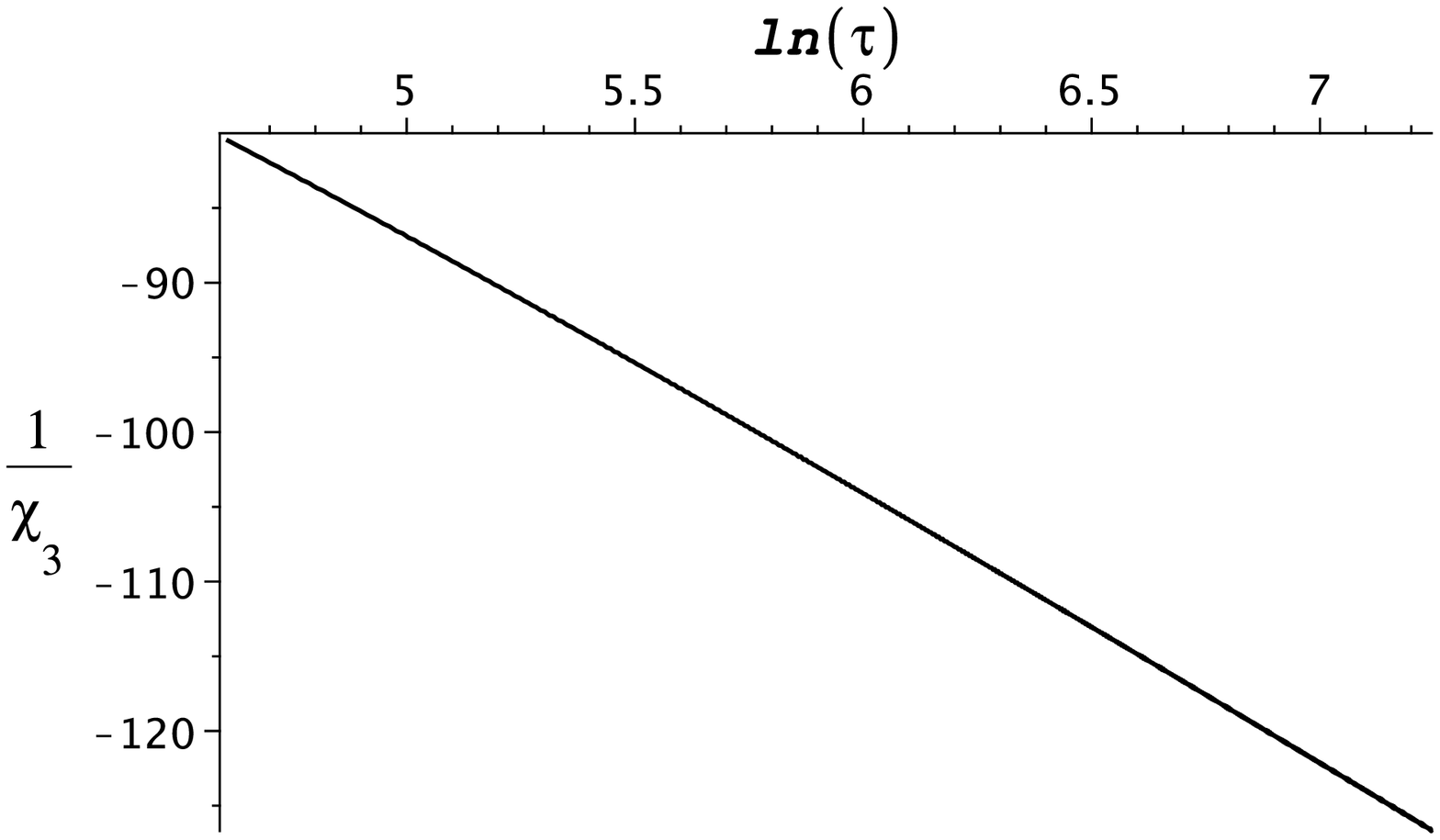}}
\caption{$1/\chi_3(\tau,r)$ as a function of $\ln(\tau)$, for $r=R_0$, for the same shell and initial conditions as in Figure 7. The linear dependence on $\ln(\tau)$ at late times is also clearly seen.}
\end{figure}

\section{Final Comments}

In this paper we have studied the evolution of MSRF initial data for the Apostolatos - Thorne model. First we analyzed in detail the relation between the configuration corresponding to the initial data and that of the assumed final static configuration, and we showed that the initial acceleration of the shell is always directed towards the static radius that corresponds to the given intrinsic conserved parameters of the shell. Then we showed that, once the appropriate properties of the solutions of the cylindrical wave equation are taken into account, there is a priori no conflict for any choice of initial MSRF data. Thus our results do not agree with those of \cite{nakao}. Next we considered the case where the problem can be analyzed in the linear approximation, and showed that the evolution is stable in all cases. The possible form of the approach to the final static configuration was also analyzed and we found that this approach is very slow, with an inverse logarithmic dependence on time at fixed radius. We also introduced a numerical computation procedure that allows us to visualize the explicit form of the evolution of the shell and of the gravitational field up to large times. The result are in agreement with the qualitative behaviour conjectured in \cite{apostol}, with an initial damped oscillatory stage, but followed by a slow approach to the static final state, as indicated by our analysis. We also include an Appendix, where we review some properties of the solutions of the cylindrical wave equation, and prove the existence of solutions with vanishing initial value for $r > R_0$, ($R_0 > 0$ some finite constant), that approach a constant value for large times. This proof is crucial for the proof of compatibility of arbitrary MSRF initial data and a final static configuration for the system.

As a final comment, we remark that for arbitrary MSRF initial data we have only shown their general compatibility with a corresponding final static configuration. Since we do not have analytic solutions, a full numerical procedure would be required to obtain a more detailed information on the actual evolution, but that is outside the scope of our paper, where a numerical procedure was used only the linearized approximation. We consider that the disagreement between our results and those of \cite{nakao} are the result their assumptions on the possible form of the function $p(\xi)$. The first and more important point is that in \cite{nako} the authors assume from the start of their use of (\ref{ap01a}) that the function $p(\xi)$ vanishes for large $\xi$, which, as we have shown, is incompatible with the possibility that the function $\psi$ goes from its initial MSRF form to the final static form. The other equally important point is that what is required to analyze the compatibility is the limit $t_+ \to \infty$ for {\em fixed} $r$, but they consider the limit $v \to \infty$, with $v=t+r$, while keeping $w=t -r$ finite. It should be clear that we would get the same results as in \cite{nakao} if we tried to compute that limit, but, as we have tried to make clear, that limit is not the relevant one for the stability analysis.

\section*{Acknowledgments}

This work was supported in part by CONICET (Argentina).

\newpage

\appendix

\section{Some properties of the solutions of the cylindrical wave equation.}

In this Appendix we analyze the properties of the solutions of the cylindrical wave equation,
\begin{equation}\label{ap00}
-\frac{\partial^2 \psi}{\partial t^2}+\frac{\partial^2 \psi}{\partial r^2}+\frac{1}{r}\frac{\partial \psi}{\partial r} = 0
\end{equation}
that are compatible with initial data such that for $t=0$, and for $r \geq R_i$, where $R_i$ is some constant, we have $\psi= \psi_i -\kappa \ln(r/R_i)$, and $\partial \psi/\partial t = 0 $. By causality, the solution for $r \geq R_i$, and for $t \geq 0$, may be written in general in the form,
\begin{equation}\label{ap00a}
\psi(r,t)= \psi_i -\kappa \ln(r/R_i) + \Phi(t,r)
\end{equation}
where $\Phi(t,r)$ is a solution of (\ref{ap00}) that is non vanishing only for $t > r - R_i \geq 0$. It may be expressed in the form,
\begin{equation}\label{ap01}
\Phi(t,r) = {\cal{P}}(t+R_i,r)
\end{equation}
where ${\cal{P}}(t,r)$ is also a solution of (\ref{ap00}), given by,
\begin{equation}\label{ap01a}
{\cal{P}}(t,r) = \int_0^{t-r}{\frac{p(\xi)}{\sqrt{(t-\xi)^2-r^2}}}d\xi
\end{equation}
and $p(\xi)$ is a function that vanishes for $\xi \leq 0$, and therefore, ${\cal{P}}(t,r)$ vanishes for $ r\geq t$.

We are interested in the behaviour of the solutions of (\ref{ap00}) given by (\ref{ap01a}), in the limit $t \to + \infty$. In particular we are interested on the existence of solutions such that, for large $t$, and finite $r$, we have,
\begin{equation}\label{ap02}
{\cal{P}}(t,r) \sim F(r)+ G(t,r)
\end{equation}
with $G(t,r) \to 0$ as $t \to \infty$. Replacing in (\ref{ap00}), in the limit $t \to \infty$ we should have,
\begin{equation}\label{ap02a}
    \frac{d^2 F}{d r^2}+\frac{1}{r}\frac{d F}{d r} = 0
\end{equation}
and, therefore, we must have,
\begin{equation}\label{ap02b}
    F(r)= A + B \ln(r)
\end{equation}

We notice that we have,
\begin{eqnarray}
\label{ap03}
  \left|{\cal{P}}(t,r)\right| & \leq &  \int_0^{t-r}{\frac{|p(\xi)|}{\sqrt{(t+r-\xi)}\sqrt{(t-r-\xi)}}}d\xi \nonumber \\
   &\leq & \frac{1}{\sqrt{2 r}}\int_0^{t-r}{\frac{|p(\xi)|}{\sqrt{(t-r-\xi)}}}d\xi
\end{eqnarray}


Let us assume first that there exist two constants, $a$ and $b$, such that , for $0 \leq \xi \leq \infty$ we have $|p(\xi)| \leq a/(\xi+b)$, that is, that $p(\xi)$ is bounded and goes to zero at least as $1/\xi$ for $\xi \to \infty$. Then we have,

\begin{eqnarray}
\label{ap04}
 \int_0^{t-r}{\frac{|p(\xi)|}{\sqrt{(t-r-\xi)}}}d\xi & \leq &  \int_0^{t-r}{\frac{a}{(\xi+b)\sqrt{(t-r-\xi)}}}d\xi \nonumber \\
   & = &  \frac{2}{\sqrt{t-r+b}}\;\;{\rm{arctanh}}\left(\frac{\sqrt{t-r}}{\sqrt{t-r+b}}\right)
\end{eqnarray}
and, since the limit $t \to \infty$ of the last expression is zero, we find that ${\cal{P}}(t,r)$ must also
vanish in that limit. Similarly, if we consider the case where $p(\xi)$ is bounded and we have $|p(\xi)| \leq a/\sqrt{\xi}$ for some finite $a$, we are lead to the bound,

\begin{eqnarray}
\label{ap05}
  \int_0^{t-r}{\frac{|p(\xi)|}{\sqrt{(t-r-\xi)}}}d\xi & \leq &  \int_0^{t-r}{\frac{a}{\sqrt{(\xi)}\sqrt{(t-r-\xi)}}}d\xi \nonumber \\
   &= &  a \pi
\end{eqnarray}
But, again, this implies that for $t \to \infty$, we have that ${\cal{P}}(t,r)$ is bounded by an expression of the form $A/\sqrt{r}$, and on account of (\ref{ap02b}), we must also have ${\cal{P}}(t,r) \to 0$ in this limit. The first example includes a large family of integrable functions that are bounded and vanish for large $\xi$, while the second includes a large family of square integrable functions that are also bounded and vanish for large $\xi$. In all these cases we have a trivial $F(r)$ as a limit.

We consider next the case where for $\xi \geq 0$ we have $p(\xi) = A$, where $A$ is constant. In this case,
\begin{equation}\label{ap06}
    {\cal{P}}(t,r) = A \Theta(t-r) \left(\ln(r)-\ln(t-\sqrt{t^2-r^2})\right)
\end{equation}
where $\Theta(x)$ is the Heaviside (step) function. For $t >> r$ we have,
\begin{equation}\label{ap07}
    {\cal{P}}(t,r) \sim -A \ln(r)+A\ln(2)+A \ln(t) -A \frac{r^2}{4 t^2}
\end{equation}
and therefore, at any fixed $r$, ${\cal{P}}$ is finite but diverges as $t \to \infty$. Notice that (\ref{ap07}) implies that there are solutions of (\ref{ap00}) that approach the form $ {\cal{P}} \sim A \ln(r) +B$ for fixed $t$, with $A$ an arbitrary constant, but where the term $B$ is time dependent and diverges for $t \to +\infty$. This result has a simple geometric interpretation. Suppose we have a solution of (\ref{ap00}) that, for fixed $r$, approaches the form $ A \ln(r) +B$, with a given fixed value of $A$, but is zero for $r > t$. Then, there will be a transition region, for $t \sim r$, where the solution changes from one regime to the other. This transition region propagates outwards with $t$ and, therefore, represents an outgoing wave, whose amplitude, because of its cylindrical nature, must decrease along with its propagation. Then, for sufficiently large $t$ we have a good approximation to the functional form of the solution by simply matching the form $A \ln(r) +B$ to zero, for $r \geq t$. But this is only possible if $B$ is of the form $B= -A \ln(t)$, which is, precisely, the form given in (\ref{ap07}). \\

Going back to (\ref{ap06}), this suggests that we consider $p(\xi)=A/\ln(\xi+b)$, with $b > 1$. We then have,
\begin{eqnarray}
\label{ap08}
 {\cal{P}}(t,r) & = & \int_0^{t-r}{\frac{A}{\ln(\xi+b) \sqrt{t-r-\xi}\sqrt{t+r-\xi}}}d\xi \nonumber \\
   &=& \int_0^1 {\frac{A}{\ln(\eta x +b) \sqrt{1-x}\sqrt{1+2 y-x}}}d x
\end{eqnarray}
where $\eta = t -r$, and $y=r/(t-r)$. Next, we may split the integral at $x=\epsilon$, with $0 <\epsilon <<1$,
\begin{eqnarray}\label{ap09}
{\cal{P}}(t,r)  & = &\int_0^{\epsilon} {\frac{A}{\ln(\eta x +b) \sqrt{1-x}\sqrt{1+2 y-x}}}d x  \nonumber \\ & & +\int_{\epsilon}^1 {\frac{A}{\ln(\eta x +b) \sqrt{1-x}\sqrt{1+2 y-x}}}d x
\end{eqnarray}
For the first integral on the right hand side of (\ref{ap09}), taking into account that $b>1$, we have,
\begin{equation}\label{ap10}
 \left|\int_0^{\epsilon} {\frac{A}{\ln(\eta x +b) \sqrt{1-x}\sqrt{1+2 y-x}}}d x \right|
    \leq  {\frac{A}{\sqrt{1-\epsilon}\sqrt{1+2 y-\epsilon}}}\int_0^{\epsilon} {\frac{1}{\ln(\eta x +b) }}d x
\end{equation}

To analyze the integral on the right hand side of (\ref{ap10}) we define the function,
\begin{equation}\label{ap11a}
 {\cal{L}}(\eta) = \int_0^{\epsilon}\frac{1}{\ln(\eta x +b)} dx
\end{equation}
and we can check that we have,
\begin{equation}\label{ap11b}
\frac{d \left(\eta{\cal{L}}\right)}{d\eta} = \frac{\epsilon}{\ln(\eta \epsilon +b)}
\end{equation}

But, this implies that for $\eta \to \infty$ we must have,
\begin{equation}\label{ap11c}
 {\cal{L}}(\eta) = \frac{\epsilon}{\ln(\eta \epsilon +b)} + {\mbox{\em{o}}}\left(\frac{1}{\ln(\eta)}\right)
\end{equation}
and, therefore, the integral in (\ref{ap10}) gives a vanishing contribution in the limit $\eta \to \infty$. Similarly,
\begin{eqnarray}
\label{ap12}
 \frac{A}{\ln(\eta  +b)}\int_{\epsilon}^1 {\frac{1}{\sqrt{1-x}\sqrt{1+2 y-x}}}d x  & \leq & \int_{\epsilon}^1 {\frac{A}{\ln(\eta x +b) \sqrt{1-x}\sqrt{1+2 y-x}}}d x
  \nonumber \\
   & \leq& \frac{A}{\ln(\eta \epsilon +b)}\int_{\epsilon}^1 {\frac{1}{\sqrt{1-x}\sqrt{1+2 y-x}}}d x
\end{eqnarray}
Evaluating the integral in the first and last terms, and replacing $y=r/\eta$,
\begin{eqnarray}
\label{ap13}
 \int_{\epsilon}^1 {\frac{1}{\sqrt{1-x}\sqrt{1+2 r/\eta-x}}}d x  & = & \ln(r/\eta)-\ln(1+r/\eta-\epsilon-\sqrt{1+2r/\eta-\epsilon}\sqrt{1-\epsilon})
  \nonumber \\
   & = & \ln(\eta)-\ln(r)+\ln(2(1-\epsilon))+{\cal{O}}(\eta^{-1})
\end{eqnarray}
Finally, collecting results from (\ref{ap10}), (\ref{ap11c}), and (\ref{ap13}), replacing in (\ref{ap12}), and taking the limit $\eta \to \infty$,
\begin{equation}\label{ap14}
 A \leq \lim_{\eta \to \infty} \int_0^1 {\frac{A}{\ln(\eta x +b) \sqrt{1-x}\sqrt{1+2 y-x}}}d x  \leq A
\end{equation}
and, therefore,
\begin{equation}\label{ap15}
 \lim_{t \to \infty} \int_0^{t-r}{\frac{A}{\ln(\xi+b) \sqrt{t-r-\xi}\sqrt{t+r-\xi}}}d\xi = A
\end{equation}

Thus we have shown that there are solutions of the wave equation (\ref{ap00}) of the form (\ref{ap01a}) that approach a given constant for any $r$ and $t \to \infty$. Notice that this will also be true for any bounded $p(\xi)$ that approaches the form $A/\ln(\xi)$ for large $\xi$. An inspection of (\ref{ap01a}) shows that the behaviour of $p(\xi)$ for small $\xi$ contains the information on the wave form emitted at small $t$.

A further relevant result from (\ref{ap14}) is that for $t>>r$, and fixed $r > R_i$ we have the approximation,
\begin{equation}\label{ap15a}
  \int_0^{t-r}{\frac{A}{\ln(\xi+b) \sqrt{t-r-\xi}\sqrt{t+r-\xi}}}d\xi \sim A \left(1-\frac{1}{\ln(t)} \ln(r)\right)
\end{equation}
This result is important because it applies to any bounded $p(\xi)$ that approaches the form $A/\ln(\xi)$ for large $\xi$, in the region $t >> r$. Notice that, at least formally, the approximation (\ref{ap15a}) is consistent with the vanishing of ${\cal{P}}(t,r)$ for $ t \sim r$. It indicates also that the approach of ${\cal{P}}(t,r)$ to the constant value $A$ is rather slow, as the coefficient of the $\ln(r)$ correction term vanishes only logarithmically with $t$. \\

We turn now to the question of the beaviour of the function $\gamma(t,r)$ for large $t$. In accordance with the previous results we assume for $\psi$ the form,
\begin{equation}\label{ap16}
    \psi(t,r)=\psi_i-\kappa \ln(r/R_i)+F(t,r)
\end{equation}
where $F(t,r)$ is an arbitrary solution of (\ref{ap00}) that vanishes for $t\leq r-R_i$, and approaches the constant value $A$ for $t \to \infty$. We then have,
\begin{eqnarray}\label{ap17}
    \frac{\partial \gamma}{\partial t} &  = & 2 r \frac{\partial \psi}{\partial r}\frac{\partial \psi}{\partial t} \nonumber \\
    & = & -2 \kappa \frac{\partial F}{\partial t} + 2 r \frac{\partial F}{\partial r}\frac{\partial F}{\partial t}
\end{eqnarray}
and also,
\begin{eqnarray}\label{ap18}
    \frac{\partial \gamma}{\partial r} &  = &  r \left[\left(\frac{\partial \psi}{\partial r}\right)^2+\left(\frac{\partial \psi}{\partial t} \right)^2 \right]\nonumber \\
    & = & \frac{\kappa^2}{r} -2 \kappa \frac{\partial F}{\partial r}+  r \left[\left(\frac{\partial F}{\partial r}\right)^2+\left(\frac{\partial F}{\partial t} \right)^2 \right]
\end{eqnarray}
Then, in the region $r \geq R_i$ we may write $\gamma$ in the form,
\begin{equation}\label{ap19}
    \gamma(t,r) = \gamma_i + \kappa^2 \ln(r/R_i) -2 \kappa F(t,r) + 2 r \int_0^t \frac{\partial F}{\partial r}\frac{\partial F}{\partial t'} dt'
\end{equation}
where we have used,
\begin{eqnarray}
\label{ap20}
  \frac{\partial}{\partial r} \left(2 r \int_0^t \frac{\partial F}{\partial r}\frac{\partial F}{\partial t'} dt'\right) &=& 2  \int_0^t \frac{\partial F}{\partial r}\frac{\partial F}{\partial t'} dt'
  +2 r \int_0^t \frac{\partial^2 F}{\partial r^2}\frac{\partial F}{\partial t'} dt'
  +2 r \int_0^t \frac{\partial F}{\partial r}\frac{\partial^2 F}{\partial r \partial t'} dt' \nonumber \\
   &=& 2 r \int_0^t \frac{\partial^2 F}{\partial t'^2}\frac{\partial F}{\partial t'} dt'
   + r \left.\frac{\partial F}{\partial r}\frac{\partial F}{\partial r}\right|_0^t \\
   & = & r \left[\left(\frac{\partial F}{\partial r}\right)^2+\left(\frac{\partial F}{\partial t} \right)^2 \right] \nonumber
\end{eqnarray}

The general form (\ref{ap19}) for $\gamma$ implies that,
\begin{equation}\label{ap21}
     \gamma(0,r) = \gamma_i + \kappa^2 \ln(r/R_i)
\end{equation}
while in the limit  $t \to \infty$ we have,
\begin{equation}\label{ap22}
   \lim_{t \to \infty} \gamma(t,r) = \gamma_i + \kappa^2 \ln(r/R_i) -2 \kappa A + 2 r \int_0^{\infty} \frac{\partial F}{\partial r}\frac{\partial F}{\partial t} dt
\end{equation}
where the last term, as can be seen taking the limit $t \to \infty$ in (\ref{ap20}), is actually a constant, independent of $r$, and, therefore, we only need to consider the limit for large $r$ to evaluate it. Clearly, the relative contributions of the last two terms in (\ref{ap22}) define the change in $\gamma$. As we shall see, (and is well known), the last term gives always a negative contribution, but, since $A$ can be negative, if the values of these terms are independent, their total contribution can be positive, or negative, or even null. To reach a definite conclusion here we need to study the general properties of the function $F(t,r)$.

Our assumption (\ref{ap16}) implies that we may write,
\begin{equation}\label{ap23}
F(t,r) = \int_0^{t'-r}{\frac{{\cal{F}}(\xi)}{\sqrt{(t'-\xi-r)(t'-\xi+r)}}}d\xi
\end{equation}
where ${\cal{F}}(\xi)$ vanishes for $\xi \leq 0$, and ${\cal{F}}(\xi) \to A/\ln(\xi)$ for large $\xi$, and $t'=t+R_i$.

We have,
\begin{equation}
\label{ap24a}
   \frac{\partial F}{\partial t} = \int_0^{t'-r}{\frac{d{\cal{F}}(\xi)}{d\xi}\frac{1}{\sqrt{(t'-\xi-r)(t'-\xi+r)}}}d\xi
\end{equation}
while,
\begin{eqnarray}
\label{ap24b}
  \frac{\partial F}{\partial r} &=&- \int_0^{t'-r}{\frac{d{\cal{F}}(\xi)}{d\xi}\frac{1}{\sqrt{(t'-\xi-r)(t'-\xi+r)}}}d\xi
  \nonumber \\
  & & - \int_0^{t'-r}{\frac{{\cal{F}}(\xi)}{\sqrt{(t'-\xi-r)(t'-\xi+r)^3}}}d\xi
\end{eqnarray}

We remark that $F(t,r)$ vanishes for $t'\leq r$, and that we are mainly interested in the limit for large $r$. We may consider now two regions for our analysis of (\ref{ap24a}) and (\ref{ap24b}). The first is for $t' \sim r$, but $r>>R_i$. In this region, since in all the integrals we have $t'+r >>\xi$, we have the approximations,
\begin{equation}
\label{ap25a}
   \frac{\partial F}{\partial t} \sim \frac{1}{\sqrt{t'+r}} \int_0^{t'-r}{\frac{d{\cal{F}}(\xi)}{d\xi}\frac{1}{\sqrt{(t'-\xi-r)}}}d\xi
\end{equation}
while,
\begin{eqnarray}
\label{ap25b}
  \frac{\partial F}{\partial r} & \sim &- \frac{1}{\sqrt{t'+r}}\int_0^{t'-r}{\frac{d{\cal{F}}(\xi)}{d\xi}\frac{1}{\sqrt{(t'-\xi-r)}}}d\xi
  \nonumber \\
  & & - \frac{1}{\sqrt{(t'+r)^3}}\int_0^{t'-r}{\frac{{\cal{F}}(\xi)}{\sqrt{(t'-\xi-r)}}}d\xi
\end{eqnarray}

Then, for large $r$, and in the region $t'\sim r$ we have that the second term in the right in (\ref{ap25b}) is small compared with the first and we have,
\begin{equation}
\label{ap26}
   \frac{\partial F}{\partial t} \sim  -\frac{\partial F}{\partial r} \sim \frac{1}{\sqrt{r}} \;\widetilde{F}(t'-r)
\end{equation}
Therefore, in the region $t' \sim r$, (\ref{ap25a}) and (\ref{ap25b}) provide the contribution from the outgoing waves to the last term in (\ref{ap22}), which, on account of the $1/\sqrt{r}$ factor, is independent of $r$ for large $r$. Notice that this contribution is always negative.

To analyze the region $t >> r $ we recall the approximation (\ref{ap15a}), valid for any ${\cal{F}}(\xi)$ that approaches the form $A/\ln(\xi)$ for large $\xi$, and, for $t >> r$, we obtain,
\begin{equation}
\label{ap27}
   \frac{\partial F}{\partial t} \sim A \frac{\ln(r)}{t \ln(t)^2} \;\;\; ; \;\;\; \frac{\partial F}{\partial r} \sim -A \frac{1}{r\ln(t)}
\end{equation}

The contribution from these terms to the last term in  (\ref{ap22}) is then of the form,
\begin{eqnarray}
\label{ap28}
  r \int_0^{\infty}{\frac{\partial F}{\partial t} \frac{\partial F}{\partial r}}dt
  & \sim & - A^2 \int_{K r}^{\infty} {\frac{\ln(r)}{t \ln(t)^3} }dt  \nonumber \\
  &  = & - A^2\frac{\ln(r)}{2 \ln(K r)^2}
\end{eqnarray}
where $K >>1$. This result implies that the contribution from the large $\xi$ behaviour of ${\cal{F}}(\xi)$ to the last term in (\ref{ap22}) vanishes in the limit $r \to \infty$, and, therefore, the total contribution of this term is independent of the limiting value $A$, and is dominated by the outgoing waves present in the region $t \sim r$.


\begin{thebibliography}{99}
\bibitem{nota1} We disregard a possible homogeneous contribution for $\xi(\tau)$, since, as can be checked explicitly,
such a contribution would be incompatible with the equations of motion in the limit $\chi_3 \to 0$. Notice that $d^2 \xi/ d\tau^2 \sim 1/\tau^2$.
\end{thebibliography}
 \end{document}